\documentclass[a4paper,11pt]{article}

\usepackage{jinstpub} 
\usepackage{lineno}
\usepackage{hyperref}
\usepackage{siunitx}
\usepackage[caption=false]{subfig}
\usepackage{amsmath}
\usepackage{booktabs}
\usepackage{multirow}
\usepackage{textcomp}
\usepackage{siunitx}

\title{\boldmath Demonstration of Gd-GEM detector design for neutron macromolecular crystallography applications}

\author[a,b,c]{D. Pfeiffer \note{Corresponding author.}}
\author[b]{F. Brunbauer}
\author[d]{V. Cristiglio}
\author[a,c,e]{R. Hall-Wilton}
\author[f,g]{M. Lupberger}
\author[h,i]{M. Mark\'{o}}
\author[b,f]{H. Muller}
\author[a]{E. Oksanen}
\author[b]{E. Oliveri}
\author[b]{L. Ropelewski}
\author[j]{A. Rusu}
\author[a,b]{J. Samarati}
\author[b,e]{L. Scharenberg}
\author[b]{M. van Stenis}
\author[b,k]{P. Thuiner}
\author[b,l]{R. Veenhof}

\affiliation[a]{European Spallation Source ESS ERIC (ESS), \\Box 176, SE-221 00 Lund, Sweden}
\affiliation[b]{European Organization for Nuclear Research (CERN), \\1211 Geneva 23, Switzerland}
\affiliation[c]{University of Milano-Bicocca, Department of Physics, \\Piazza della Scienza 3, 20126 Milan, Italy}
\affiliation[d]{Institut Laue-Langevin, \\71 avenue des Martyrs CS 20156, 38042 Grenoble, France}

\affiliation[e]{Fondazione Bruno Kessler, Sensors \& Devices Centre, \\via Sommarive 18, 38123 Trento, Italy}
\affiliation[f]{Physikalisches Institut, University of Bonn, \\Nu{\ss}allee 12, 53115 Bonn, Germany}
\affiliation[g]{Helmholtz-Institut für Strahlen- und Kernphysik, University of Bonn, \\Nu{\ss}allee 14--16, 53115 Bonn, Germany}
\affiliation[h]{Neutron Spectroscopy Department,Institute for Energy Security and Environmental Safety,Centre for Energy Research\\29-33. Konkoly-Thege Mikl\'{o}s \'{u}t, Budapest, 1121, Hungary}
\affiliation[i]{Budapest Neutron Centre\\29-33. Konkoly-Thege Mikl\'{o}s \'{u}t, Budapest, 1121, Hungary}

\affiliation[j]{SRS Technology, 30 Promenade des Artisans, 1217 Meyrin, Switzerland}
\affiliation[k]{Vienna University of Technology,\\1040 Vienna, Austria}
\affiliation[l]{Bursa Uluda{\u g} University, \\G{\"o}r{\"u}kle Kampusu, 16059 Ni{\"u}fer/Bursa, Turkey}

\emailAdd{dorothea.pfeiffer@ess.eu}

\abstract{The European Spallation Source (ESS) in Lund, Sweden will become the world's most powerful thermal neutron source. The Macromolecular Diffractometer (NMX) at the ESS requires three 51.2 x 51.2~cm$^{2}$ detectors with reasonable detection efficiency,  sub-mm spatial resolution, a narrow point-spread function (PSF), and good time resolution. This work presents measurements with the improved version of the NMX detector prototype consisting of a Triple-GEM (Gas Electron Multiplier) detector with a natural Gd converter and a low material budget readout. The detector was successfully tested at the neutron reactor of the Budapest Neutron Centre (BNC) and the D16 instrument at the Institut Laue-Langevin (ILL) in Grenoble. The measurements with Cadmium and Gadolinium masks in Budapest demonstrate that the point-spread function of the detector lacks long tails that could impede the measurement of diffraction spot intensities. On the D16 instrument at ILL, diffraction spots from Triose phosphate isomerase w/ 2-phosphoglycolate (PGA) inhibitor were measured both in the MILAND Helium-3 detector and the Gd-GEM. The comparison between the two detectors shows a similar point-spread function in both detectors, and the expected efficiency ratio compared to the Helium-3 detector. Both measurements together thus give good indications that the  Gd-GEM detector fits the requirements for the NMX instrument at ESS.}


\begin{document}
\maketitle
\flushbottom

\section{Introduction}
\label{introduction}
The European Spallation Source (ESS)~\cite{ESS} in Lund, Sweden is foreseen to start the scientific user programme in 2025, and the construction phase will be complete by 2027. It will become the world's most powerful thermal neutron source with a significantly higher brilliance than at existing reactor sources such as the Institut Laue-Langevin (ILL)~\cite{ILL} or other spallation sources such as the Spallation Neutron Source (SNS)~\cite{SNS} and the Japan Proton Accelerator Research Complex (J-PARC)~\cite{J-PARC}. 15 neutron instruments constitute the initial instrument suite at ESS~\cite{Peggs2013, Andersen2020}. For each of the instruments efficient thermal neutron detectors are a crucial component~\cite{Kirstein2014}. The Macromolecular Diffractometer NMX~\cite{Andersen2020} requires three 51.2 x 51.2~cm$^{2}$ detectors with reasonable detection efficiency (around 25$\%$ for cold neutrons in the center of the wavelength range from 1.8 to 10 \AA) and sub-mm spatial resolution~\cite{NMX2020}. While spatial resolution is often used as a requirement for macromolecular crystallography detectors, it is worth noting that the shape and width of the point-spread function have an even larger effect on the ability to reliably measure the intensities of diffraction spots. Neutron image plates~\cite{Niimura1994} that are used at reactor sources macromolecular crystallography instruments have a sufficient spatial resolution of 200~$\mu$m, but lack the time resolution needed at spallation source instruments. Scintillation-based detectors~\cite{Hosoya2009, Coates2010} offer time resolution, but are currently limited to about 1~mm spatial resolution. 

For spallation source instruments the combination of solid neutron converters with Micro Pattern Gaseous Detectors (MPGDs)~\cite{Titov2013,Guerard2012} is a promising option to achieve a similar spatial resolution to the neutron image plate with sufficient time resolution, high-rate capabilities and a reasonable neutron detection efficiency. A time-resolved detector is necessary to be able to carry out a Time-of-Flight analysis of the neutrons in the diffraction spots, to disentangle the spatial overlap of reflections that occur at the same location, but at different wavelengths. The nanosecond time resolution of the detector and readout system~\cite{Scharenberg2020} is far better than the required microsecond resolution. The rate capabilities of Triple-GEM detectors~\cite{Thuiner2016} and the readout system~\cite{Pfeiffer2022} together support the requirements for NMX, which are estimated to be less than one MHz per cm$^{2}$. To obtain a good spatial resolution, the $\mu$TPC data analysis technique, based on the principle of the Time Projection Chamber (TPC), was developed. Charge produced close to the neutron converter has to drift to the amplification stage of the detector, whereas charge produced close to the first GEM is directly amplified. The $\mu$TPC method retains therefore the x and y-strip with the largest time as start of the ionisation track and thus as position where the neutron interacted in the neutron converter. The method has first been successfully applied to MPGDs with $^{10}$B$_4$C converters~\cite{Hoglund2012}. A position resolution with a $\sigma$ better than 200~$\mu$m was obtained by determining the starting point of the neutron-induced ionization track~\cite{Pfeiffer2015}. To improve the low detection efficiency (less than 5$\%$ at 1.8 \AA), the $^{10}$B$_4$C converter has been replaced with a more efficient gadolinium converter. 

A prototype of the NMX detector~\cite{Pfeiffer2016} consisting of a Triple-GEM detector~\cite{Altunbas2002} with Gd converter has first been tested in 2016 at the R2D2 beamline at the JEEP II research reactor at the Institute for Energy Technology (IFE) in Kjeller/Norway~\cite{IFE}. With the natural Gd converter, a position resolution with a $\sigma$ better than 400~$\mu$m was measured. The properties of natural and isotope-enriched Gd ($^{157}$Gd) neutron converters have been studied in detail in work package 4 of the European Horizon2020 grant BrightnESS~\cite{BrightnESS2015}. This work presents measurement with the improved version of the NMX detector prototype at the neutron reactor of the Budapest Neutron Centre (BNC)~\cite{BNC} and at the D16 instrument at the Institut Laue-Langevin (ILL)~\cite{ILL} in Grenoble in the year 2018. The purpose of these measurements was to determine the spatial resolution of the detector and the point-spread function (PSF) with realistic samples. As it is a well-established practice in crystallography to approximate the PSF with a Gaussian~\cite{Dials2018}, the paper subsequently uses Gaussian fits for this purpose.

\section{The NMX detector prototype}
\label{NMX detector}
\begin{figure}[htbp]
\centering
\subfloat[Schematic view of Triple-GEM detector\label{fig: Detector_Scheme}]{
\includegraphics[width=.56\textwidth]{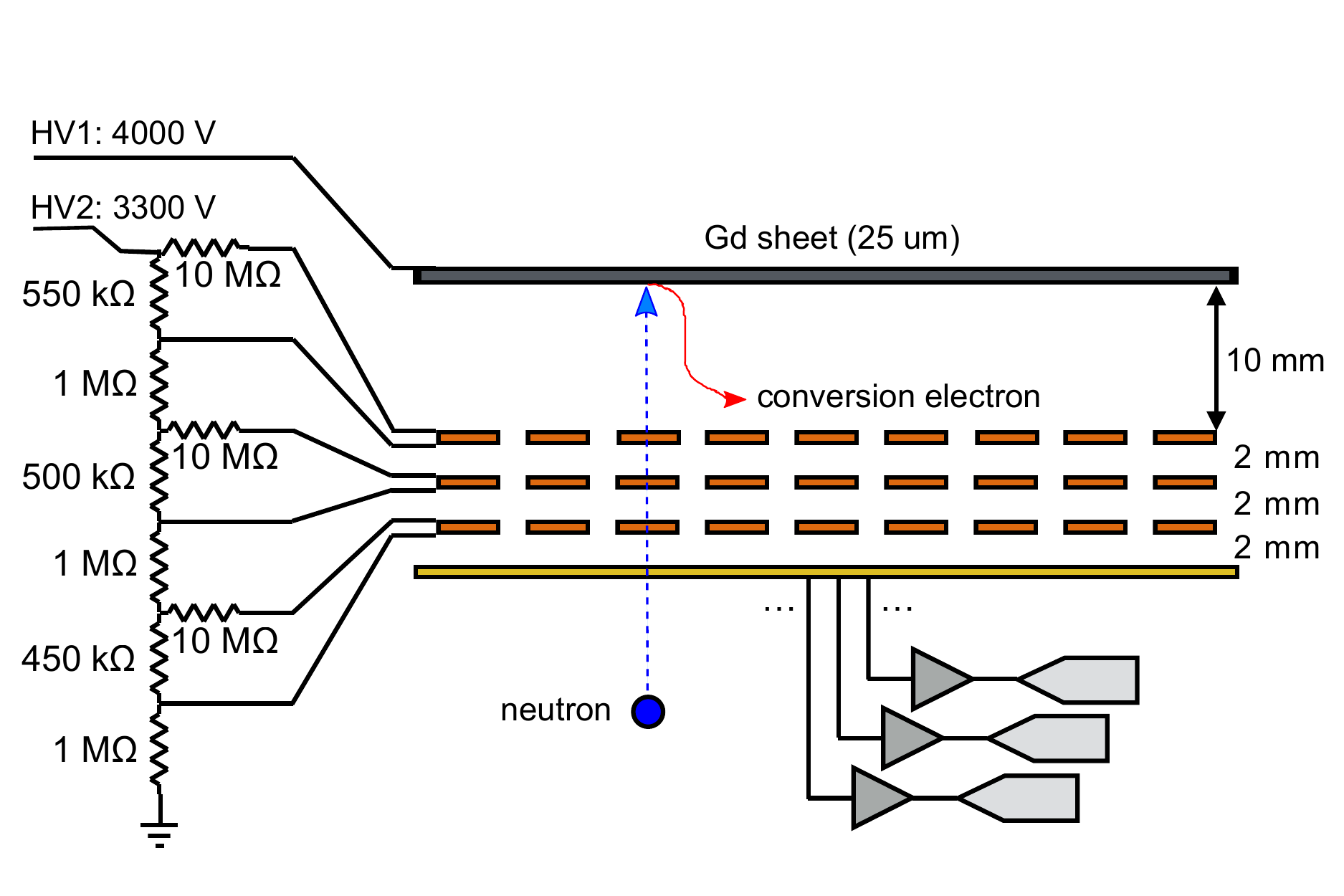}%
}
\subfloat[Gd-GEM 10~cm x 10~cm prototype\label{fig: Detector_Photo}]{
\includegraphics[width=.43\textwidth]{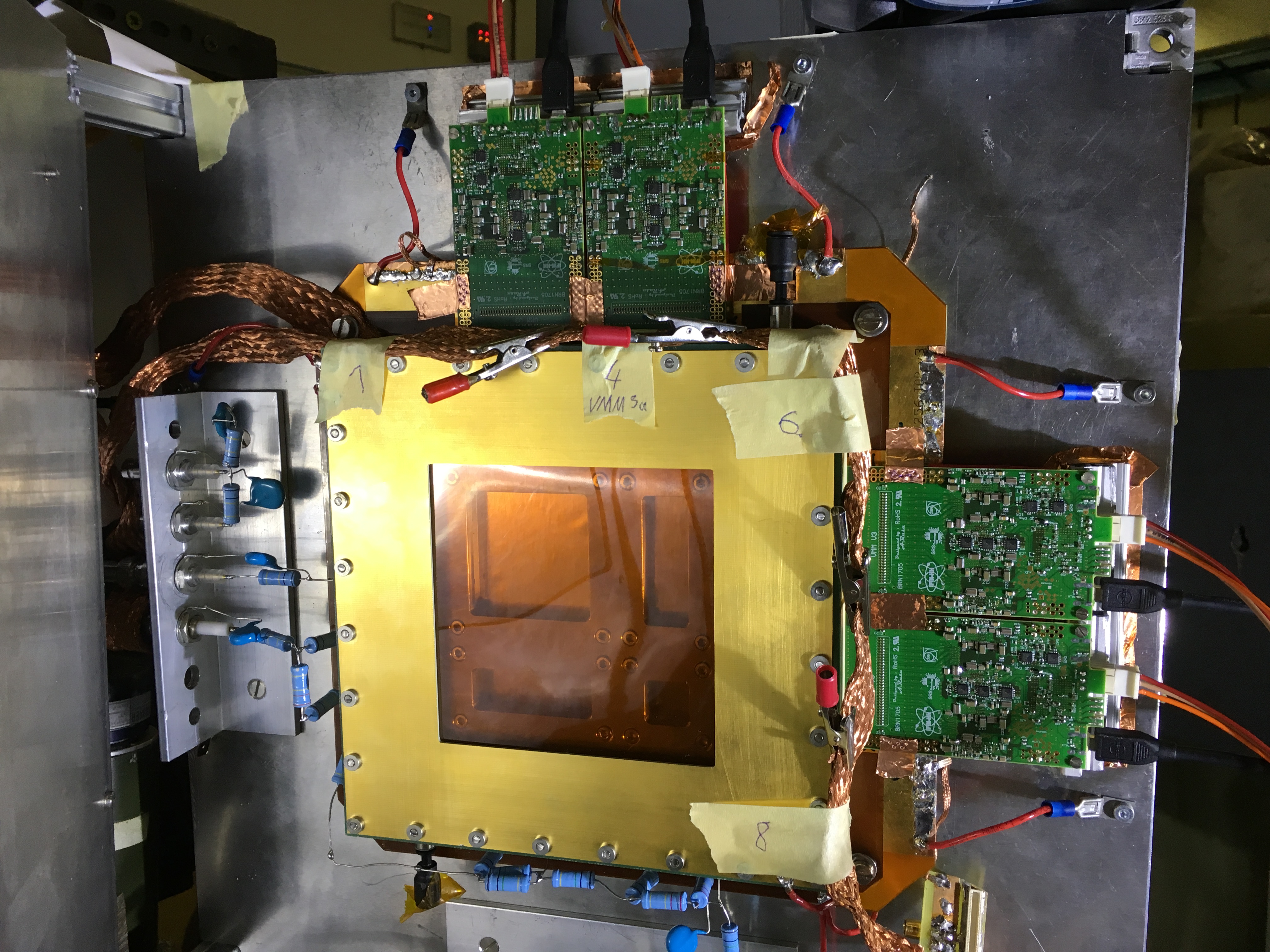}%
}
\caption{Gd-GEM neutron detector}
\label{fig:Detector}
\end{figure}

A schematic drawing of the NMX detector is displayed in figure \ref{fig: Detector_Scheme}. The NMX detector is a triple GEM detector~\cite{Altunbas2002} with a low material budget x/y strip readout. In contrast to the traditional x/y strip readout that contains an FR4 base~\cite{Bressan1999}, the low material budget readout uses a 100~$\mu$m thick Kapton foil to reduce the neutron scattering from about 20~$\%$ to about 5~$\%$. The active area of the final NMX detector is 51.2 cm x 51.2 cm, with 1280 strips in x and y. The strip pitch is 400 $\mu$m. In the present article though, a small prototype with 10 cm x 10 cm active area (256 strips in x and y) has been tested. The detector was flushed with Ar/CO$_2$ 70/30 mixture at 5~l/h at room temperature and atmospheric pressure. The conversion volume or drift space was 10~mm long, and as cathode and neutron converter served a 25~$\mu$m Gadolinium foil. A drift field of 700 V/cm was chosen to avoid electron attachment to electronegative impurities and the subsequent loss of primary ionization electrons, while keeping the drift velocity smaller than 2.0~cm/$\mu$ at atmospheric pressure. The three GEM foils~\cite{Sauli1997} in the detector are mounted at a distance of 2~mm from each other and powered with a resistor chain. The values of the resistors are stated in figure \ref{fig: Detector_Scheme}. The detector was operated at an effective gain of about 5000. This small prototype is read out with two RD51 VMM3a hybrids each in x and y directions (figure \ref{fig: Detector_Photo}). Two VMM3a ASICs~\cite{DeGeronimo2012,Iakovidis2020} with 64 channels per chip are mounted on a RD51 VMM3a hybrid~\cite{Lupberger2018, Pfeiffer2022}, resulting in 128 channels per hybrid. The Scalable Readout System (SRS)~\cite{Martoiu2013} serves as readout system to acquire the data from the RD51~\cite{RD51} VMM3a hybrids.

The detector is operated in backwards mode. Neutrons traverse the detector readout and the three GEM foils, and then impinge on the natural Gd cathode. Natural Gd contains 14.8 $\%$ of $^{155}$Gd and 15.7 $\%$ of $^{157}$Gd. Both of these isotopes have a very high neutron capture cross section~\cite{Harms1974}. After the neutron capture, gadolinium releases prompt gamma particles with an energy of up to 9 MeV and conversion electrons with energies ranging from 29 keV to 250 keV. The signals resulting from these conversion electrons stopped in the drift volume, are used to determine the position of the neutron impact on the cathode with the help of the uTPC method~\cite{Pfeiffer2015, Pfeiffer2016}.

\section{Test beam in Budapest}
\label{Budapest}
\begin{figure}[htbp]
\centering
\subfloat[Cd mask, 1.0~mm holes\label{fig: Cd_mask_1p0_photo}]{
\includegraphics[width=.475\textwidth]{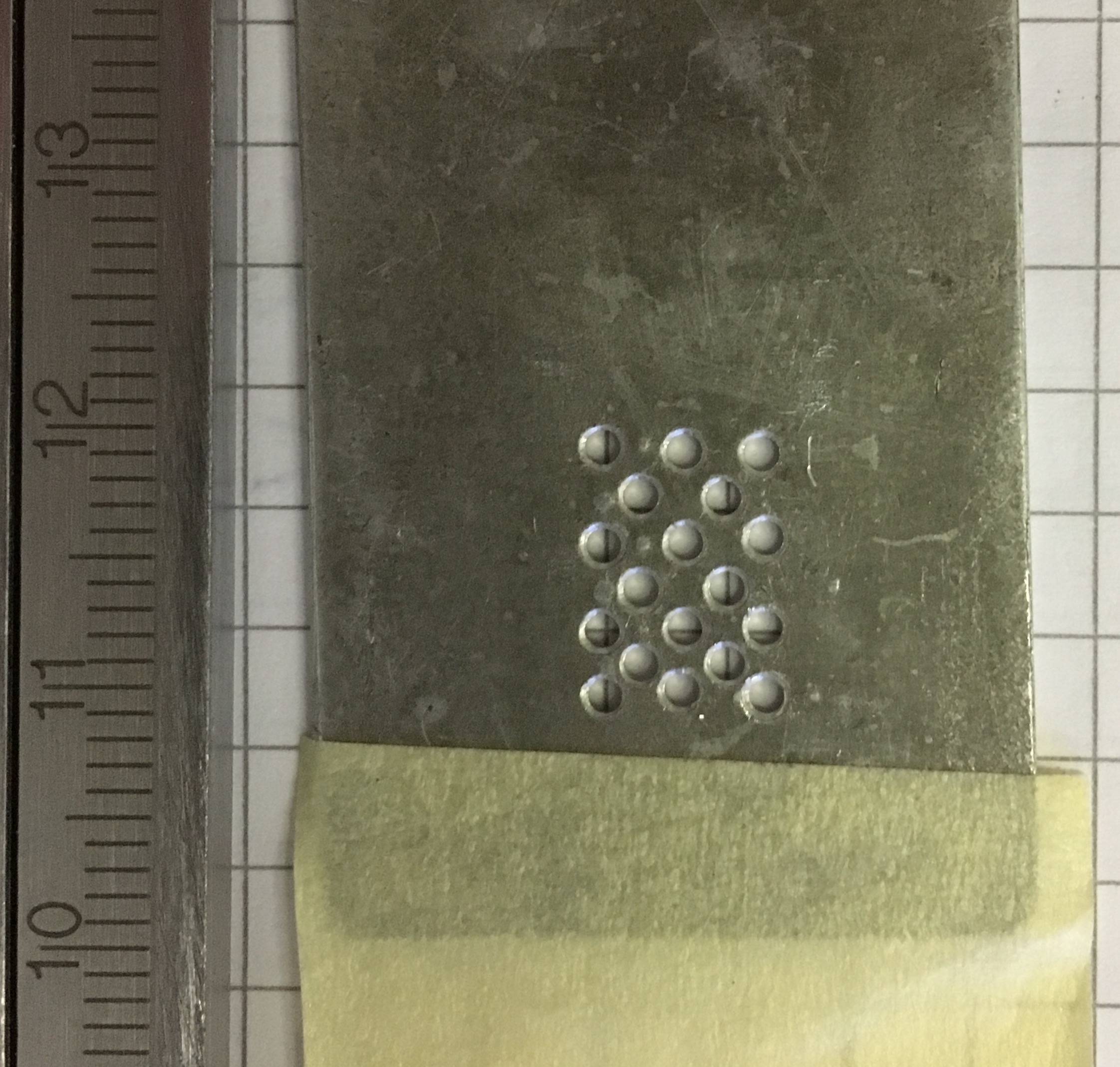}%
}
\subfloat[Measurement with Cd mask\label{fig: Cd_mask_1p0_measurement}]{
\includegraphics[width=.525\textwidth]{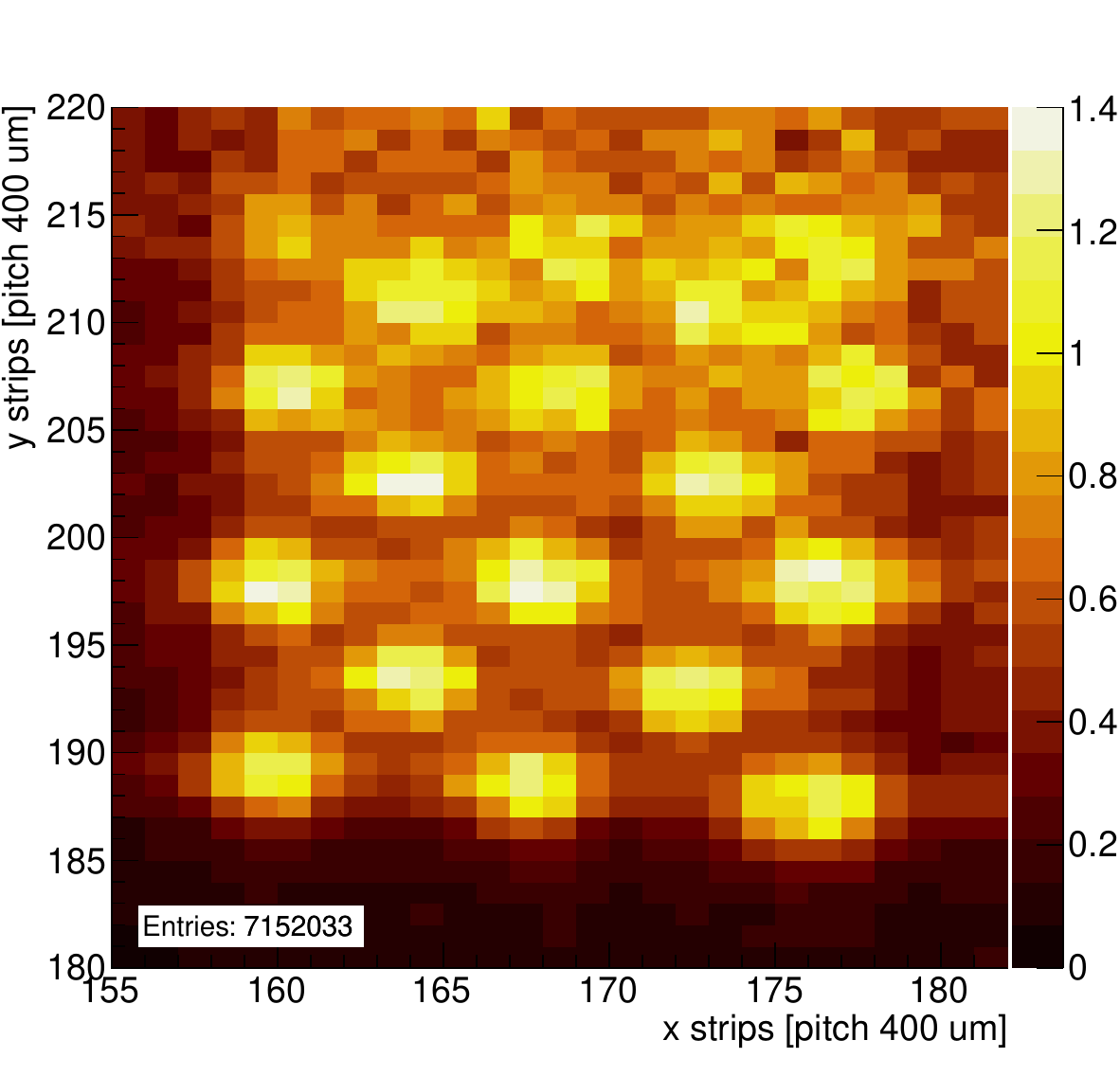}%
}
\caption{Recorded pattern of a Cd mask with circular holes of 1.0 mm diameter. Figure a) show a picture of the mask that was installed on the detector. The holes have a minimum separation (centre to centre) of about 2 mm horizontally, 1.6 mm vertically, and 1.3 mm diagonally. Figure b) shows the beam intensity recorded with the mask normalized to the beam intensity without mask. The beam was focused on the lowest three rows of the mask. The holes there can be clearly resolved. }
\label{fig:Cd_mask}
\end{figure}

The detector prototype was tested in 2018 at the neutron reactor of the Budapest Neutron Centre (BNC). A monochromatic neutron beam at the ATHOS~\cite{ATHOS2004} triple-axis spectrometer with a wavelength of 3.5 \AA~was used for the measurements. Figure \ref{fig: Cd_mask_1p0_photo} shows a Cadmium mask with circular holes of 1.0 mm diameter. The holes of the mask have a minimum separation (centre to centre) of about 2 mm horizontally, 1.6 mm vertically, and 1.3 mm diagonally. To record this hole pattern with the detector, the beam was collimated to 1.1 cm in x direction and 3.3 cm in y direction. First data were acquired with the collimated beam directly hitting the detector without mask for 4 minutes. Then the mask was taped to the readout of the detector, and data were acquired for 84 minutes. To get rid of detector and electronics-specific inhomogeneities, the mask data were normalized with the direct beam data. Figure \ref{fig: Cd_mask_1p0_measurement} shows the recorded image of the mask. The collimated beam was directed at the lowest three rows of the mask. The holes there can be clearly resolved.

\begin{figure}[htbp]
\centering
\subfloat[Measurement with Gd-foil mask with 1 mm hole\label{fig: single_hole}]{
\includegraphics[width=.45\textwidth]{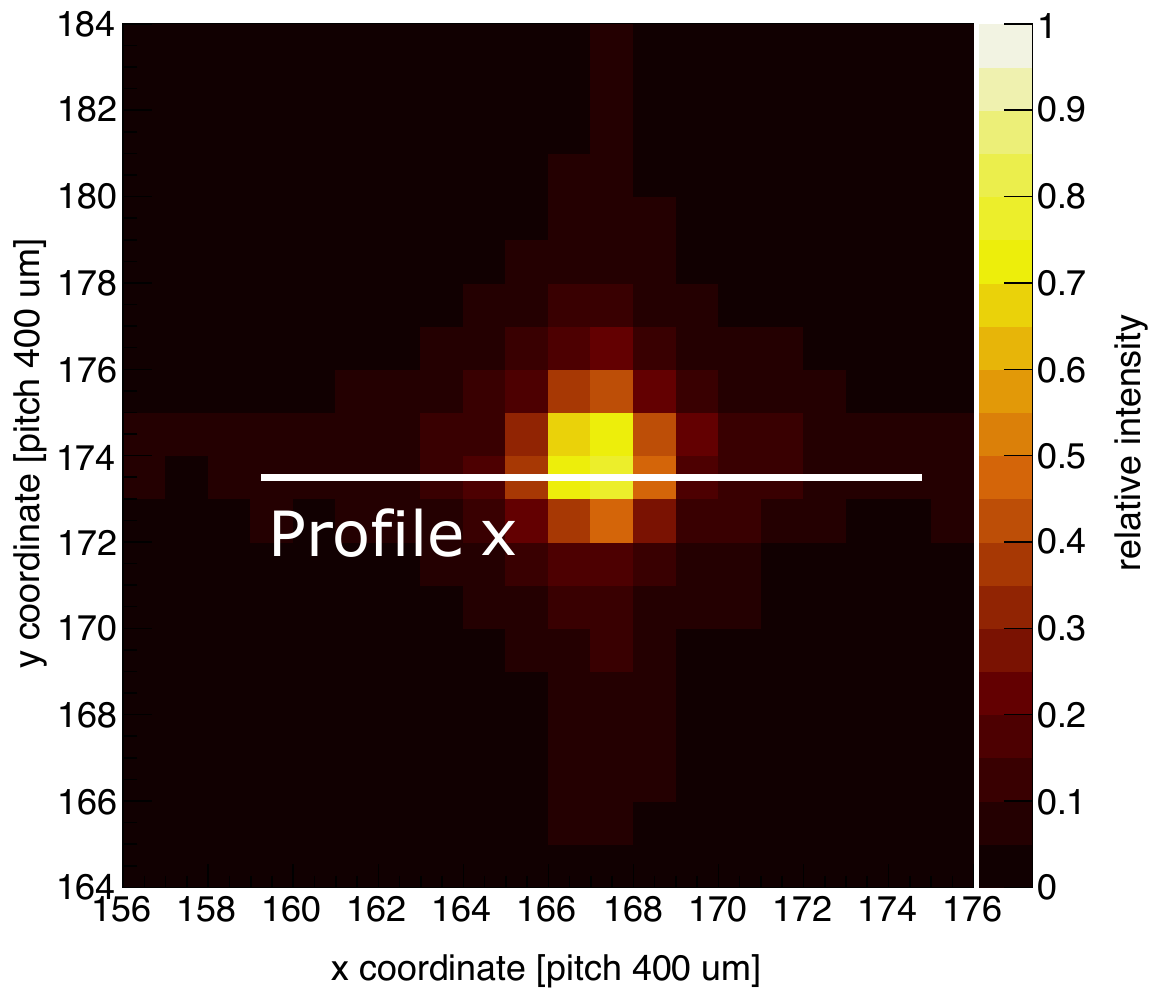}%
}
\subfloat[Profile with 1.4 mm FWHM\label{fig: single_hole_profile}]{
\includegraphics[width=.55\textwidth]{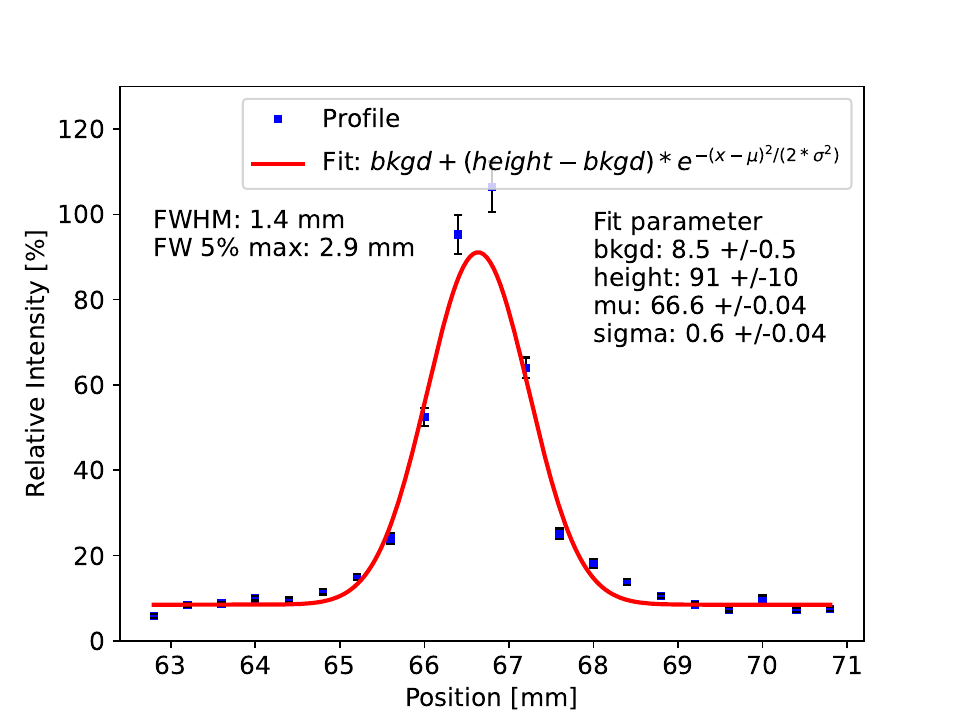}%
}
\caption{Measurement with Gd-foil with 1 mm hole in front of detector readout. The Gaussian fitted to the profile has a FWHM of 1.4 mm. The binning corresponds to the 400 $\mu$m strip pitch of the detector readout.}
\label{fig:Single_hole}
\end{figure}

To further study the spatial resolution of the detector, the centre part of the readout of the detector was covered with a 25 um thick Gd foil with a single 1 mm diameter hole. Data were acquired for 135 minutes. Figure \ref{fig: single_hole_profile} shows the measurement. As in the case of the Cadmium mask data, the single-hole data has been normalized with the direct beam data.

The Gaussian ($\sigma$ 600 $\mu$m) fit of the profile in x direction is displayed in figure \ref{fig: single_hole_profile}. The error bars indicate the statistical uncertainty on the counts. The Full-Width-Half-Maximum (FWHM) is 1.4 mm, and the full width at 5 \% of the height is 2.9 mm. 

Figure~\ref{fig: single_hole_profile} and the corresponding width suggests that the point-spread function of the detector lacks long tails that could impede the measurement of diffraction spot intensities. While a hole in a mask attached to the detector surface is not entirely comparable to a diffraction spot, it is a good measure of the point-spread function of such a spot due to detector instrumental effects. This result indicates that the detector is suitable for measuring the diffraction spot intensities in crowded diffraction patterns.

\section{Test beam at ILL Grenoble}
\label{ILL}
\begin{figure}[htbp]
\centering
\subfloat[D16 beamline with detector\label{fig: Detectors_Grenoble}]{
\includegraphics[width=.65\textwidth]{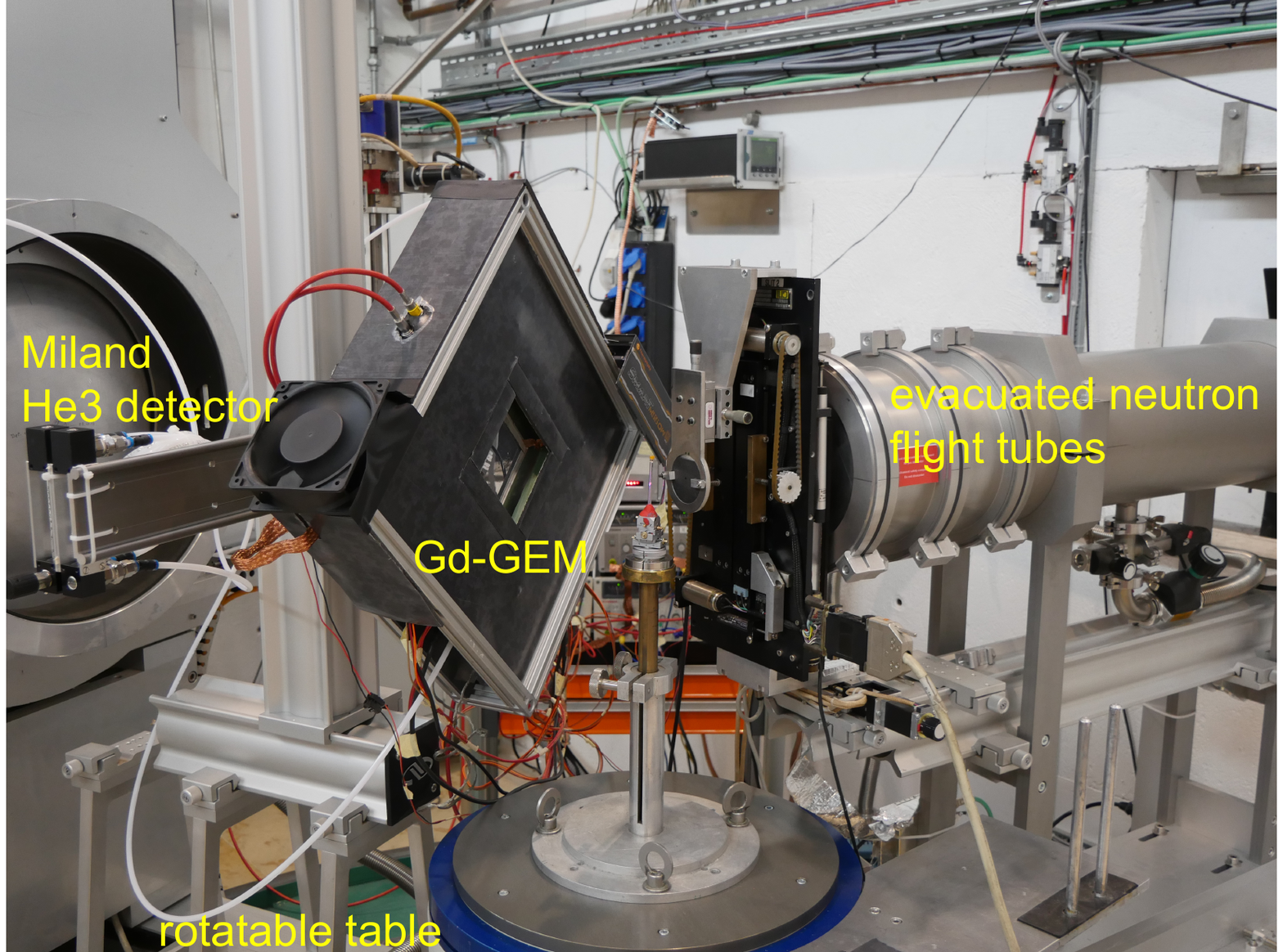}%
}
\subfloat[B$_{4}$C diaphragm and sample holder\label{fig: Diaphragm_sample}]{
\includegraphics[width=.35\textwidth]{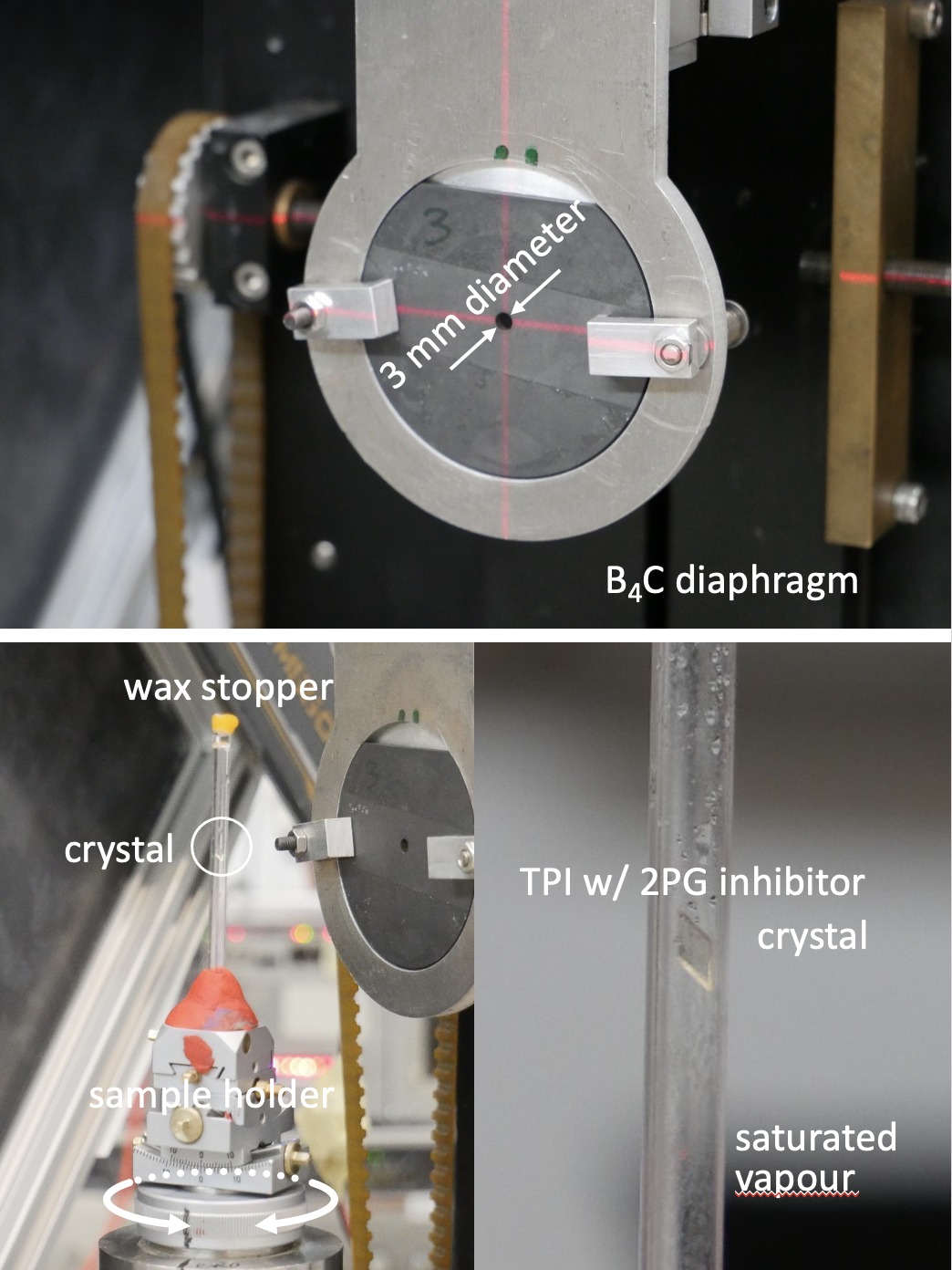}%
}
\caption{Beamline of D16 instrument at ILL in Grenoble. The Gd-GEM detector can be seen in the center of \ref{fig: Detectors_Grenoble}, whereas the MILAND detector is the circular chamber visible on the outer left. The neutron beam traverses a 3 mm $B_{4}C$ diaphragm and hits the PGA inhibitor sample. The Gd-GEM detector is mounted at a distance of 196 mm from the sample.}
\label{fig:Beamline_Grenoble}
\end{figure}

Also in 2018 the detector has been tested against the MILAND detector (Helium-3 based wire chamber with a wire plane in x and y direction) of the D16~\cite{D16} instrument at the Institut Laue-Langevin (ILL)~\cite{ILL} in Grenoble. Monochromatic neutrons with a wavelength of 4.51 \AA~were used for the measurements. The beamline of the D16 instrument is displayed in figure \ref{fig: Detectors_Grenoble}. In the foreground of the picture, the sample and the Gd-GEM detector are visible. On the left-hand side the circular vessel of the D16 $^{3}He$ wire chamber can be seen in its retracted position.

The neutron beam traverses a 3 mm natural B$_{4}$C diaphragm and hits the sample (Triose phosphate isomerase w/ 2-phosphoglycolate (PGA) inhibitor)~\cite{TIM2019} (figure \ref{fig: Diaphragm_sample}). The Gd-GEM detector in its shielding box was mounted at a distance of 140 mm from the sample. The cathode of the detector, where the impinging neutrons are converted into conversion electrons, is 56 mm away from the beginning of the detector shielding box. The cathode of the detector was thus at a distance of 196 mm from the sample.

\begin{figure}[htbp]
\centering
\subfloat[Detector rotation $\gamma$\label{fig: Detector_orientation}]{
\includegraphics[width=.50\textwidth]{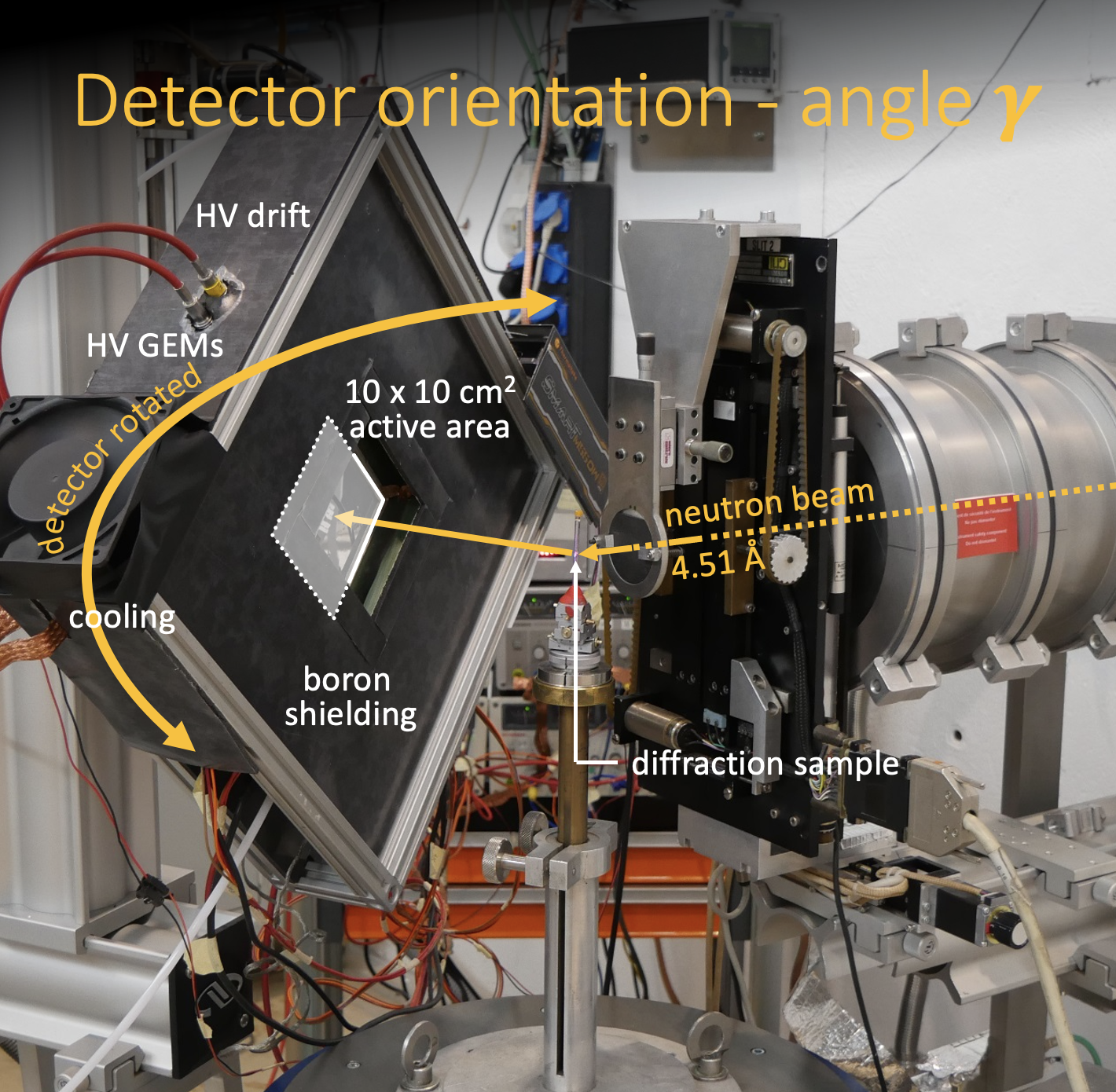}%
}
\subfloat[Sample orientation $\omega$\label{fig: Sample_orientation}]{
\includegraphics[width=.50\textwidth]{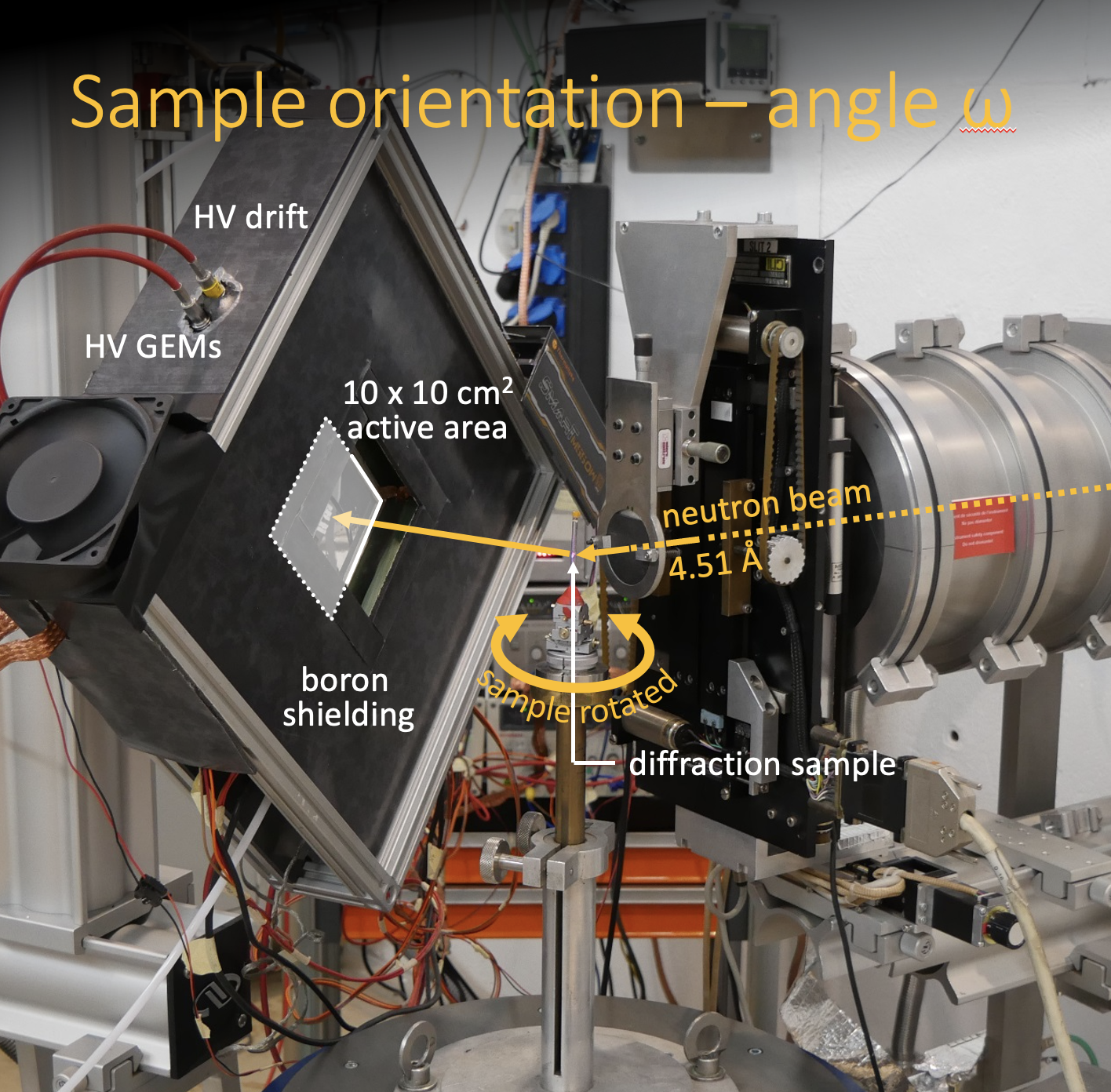}%
}
\caption{The detector and sample orientation (angles $\gamma$ and $\omega$) can be adjusted from 0 to 360 degrees. }
\label{fig:orientation_detector}
\end{figure}

The Gd-GEM and the MILAND detector were both fixed to a rotatable table seen at the bottom left corner. To take data with the MILAND detector, the Gd-GEM and its support have to be removed, since they block the neutrons coming from the sample. Subsequently, the MILAND detector is then moved forward towards the sample into its data acquisition position (332 mm from the sample). The angle $\gamma$ indicates the rotation of the detectors with respect to the end of the beamline as indicated in figure \ref{fig: Detector_orientation}. The sample itself is mounted on a goniometer, and can also be rotated around its axis. The angle $\omega$ thus indicates the rotation of the sample, as explained in figure \ref{fig: Sample_orientation}.

The setup that was used at the D16 instrument was not optimized to study the unit cell of PGA inhibitor. The Gd-GEM detector could not be installed at the same location as the MILAND detector. Since the Gd-GEM with an active area of 10 cm x 10 cm is also nine times smaller than the MILAND detector, the Gd-GEM was installed closer to the sample. To summarize, it is not obvious which reflection measured on the MILAND detector belongs to which reflection on the Gd-GEM. Considering the different solid angles covered by the two detectors, geometrical calculations were needed to determine the matching reflection pairs. The details of the geometrical calculations can be found in the annex (chapter \ref{Annex}).

\begin{figure}[htbp]
\centering
\subfloat[$\omega = 5\si{\degree}$\label{fig: He3_GdGem_omega5_gamma60}]{
\includegraphics[width=.50\textwidth]{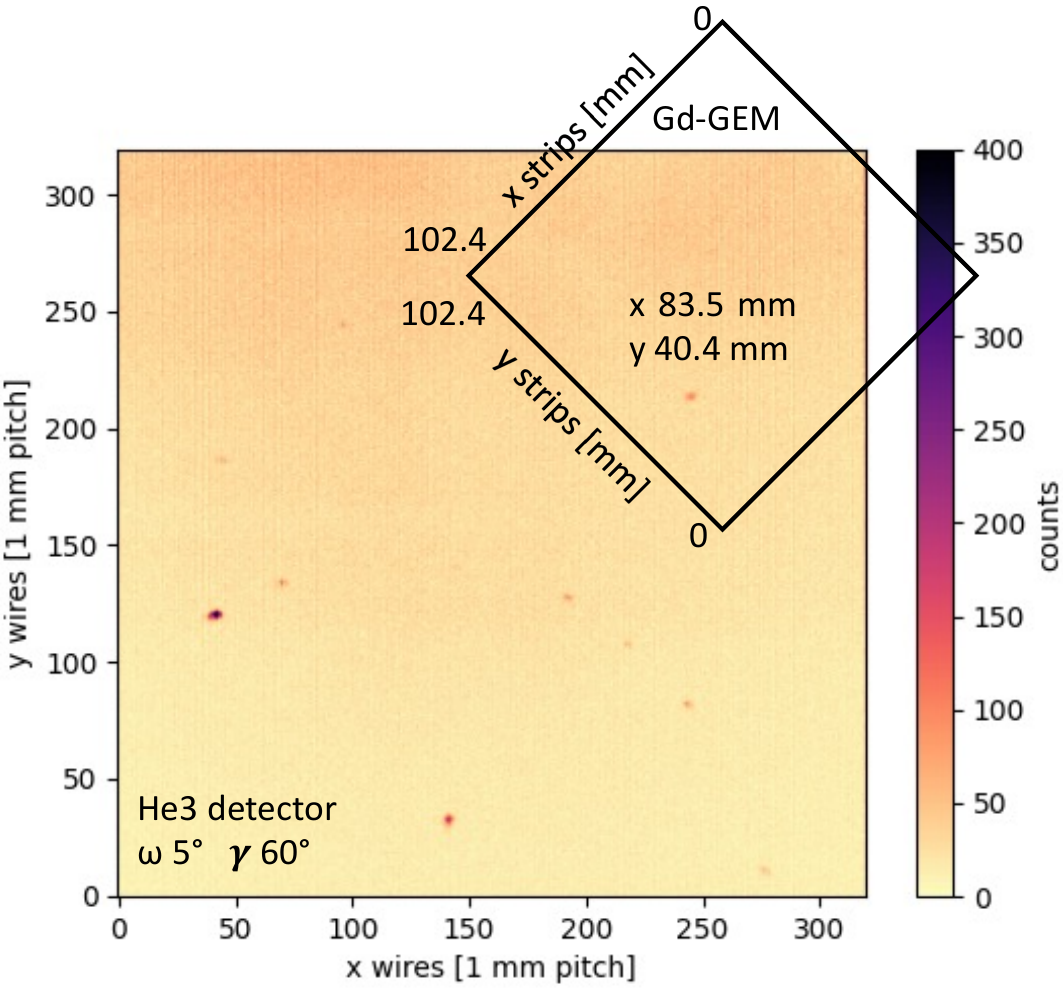}%
}
\subfloat[$\omega = 55\si{\degree}$\label{fig: He3_GdGem_omega55_gamma60}]{
\includegraphics[width=.50\textwidth]{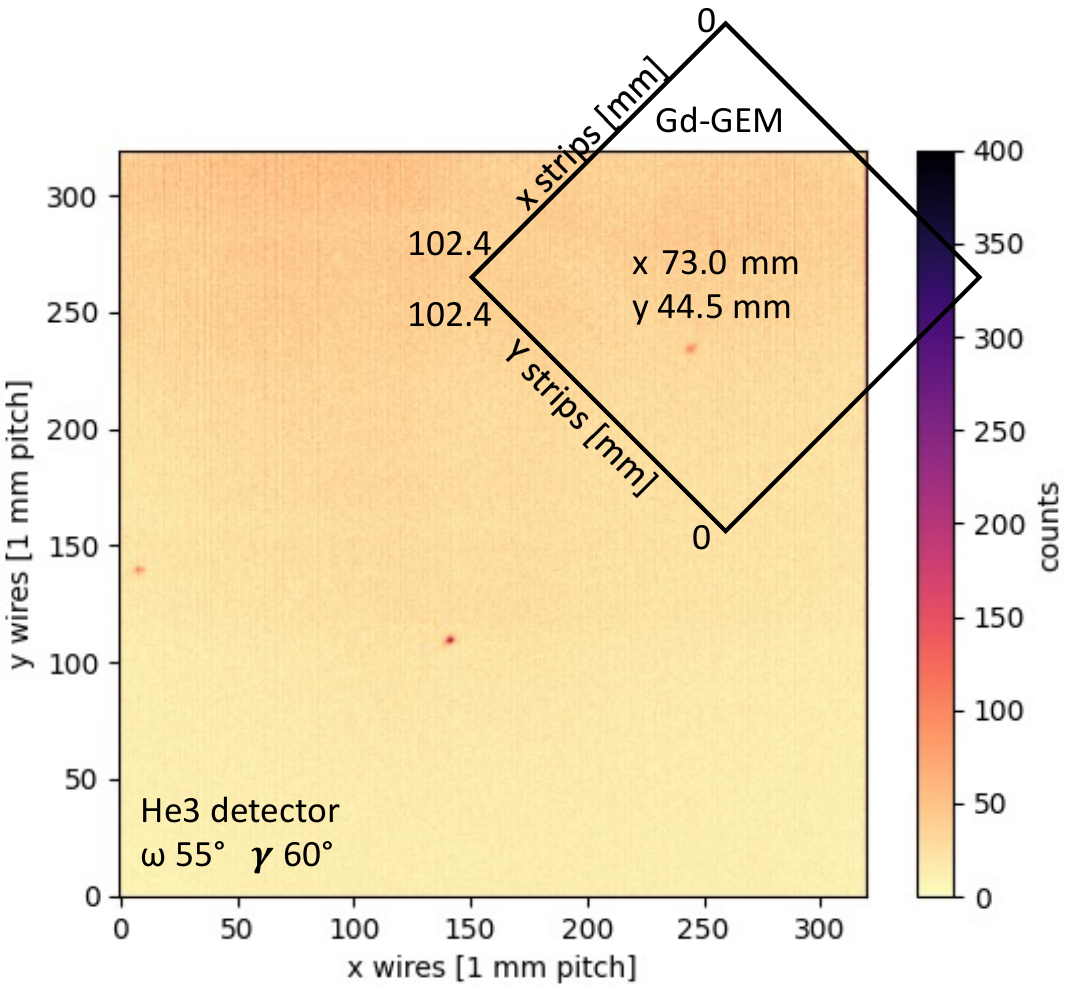}%
}
\caption{Measurement of the diffraction spots at $\omega = 5\si{\degree}$ and $\omega = 55\si{\degree}$ in the MILAND detector. The detector rotation was $\gamma = 60\si{\degree}$. The MILAND detector acquired in both measurements data for 300~s. Whereas in the MILAND detector several diffraction spots are visible (8 spots at the right and 3 spots at the left), the smaller Gd-GEM detector recorded only 1 spot in both measurements. The centre of the Gd-GEM detector has been misaligned with respect to the centre of the MILAND detector. }
\label{fig:comparison_He3_GdGEM}
\end{figure}

Figure~\ref{fig:comparison_He3_GdGEM} compares at the detector rotation $\gamma = 60\si{\degree}$ the measurements in the MILAND detector with the Gd-GEM detector. The sample orientation was changed from $\omega = 5\si{\degree}$ in figure~\ref{fig: He3_GdGem_omega5_gamma60} to $\omega = 55\si{\degree}$ in figure~\ref{fig: He3_GdGem_omega55_gamma60}. Whereas in the MILAND detector several diffraction spots are visible (8 spots at the right and 3 spots at the left), the Gd-GEM detector recorded only 1 spot in both measurements due to its smaller physical size. It is apparent that the centre of the Gd-GEM detector has been misaligned with respect to the centre of the MILAND detector.

\begin{table}[h!]
\centering
\begin{center}
\begin{tabular}{||c | c | c ||} 
 \hline
 & \multicolumn{2}{c||}{Measured/Calculated} \\
 Detector & Spot distance & Detector rotation \\ [0.5ex] 
 \hline\hline
 Gd-GEM & 11.46(39)~mm & 3.35(12) \si{\degree} \\
 MILAND $^{3}He$ & 20.63(17)~mm & 3.56(3) \si{\degree} \\ 
\hline
\end{tabular}
\end{center}
\caption{Measurements at $\gamma = 60\si{\degree}$ detector rotation and a sample rotation of $\omega = 5\si{\degree}$ and $\omega = 55\si{\degree}$. The table shows the distance between spots in the MILAND and Gd-GEM detector, and the equivalent detector rotation. The equal equivalent detector rotation shows that the spots belong to the same reflection in both detectors.}
\label{table:detector_rotation}
\end{table}

\begin{figure}[htbp]
\centering
\subfloat[MILAND detector\label{fig: fit_1D_spot_He3}]{
\includegraphics[width=.47\textwidth]{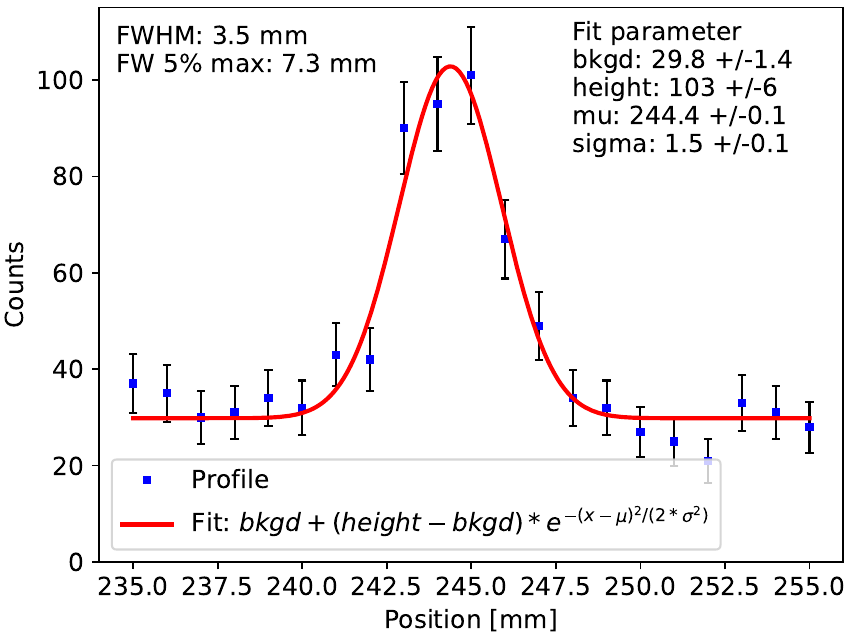}%
}
\subfloat[Gd-GEM detector\label{fig: fit_1D_spot_Gd-GEM}]{
\includegraphics[width=.53\textwidth]{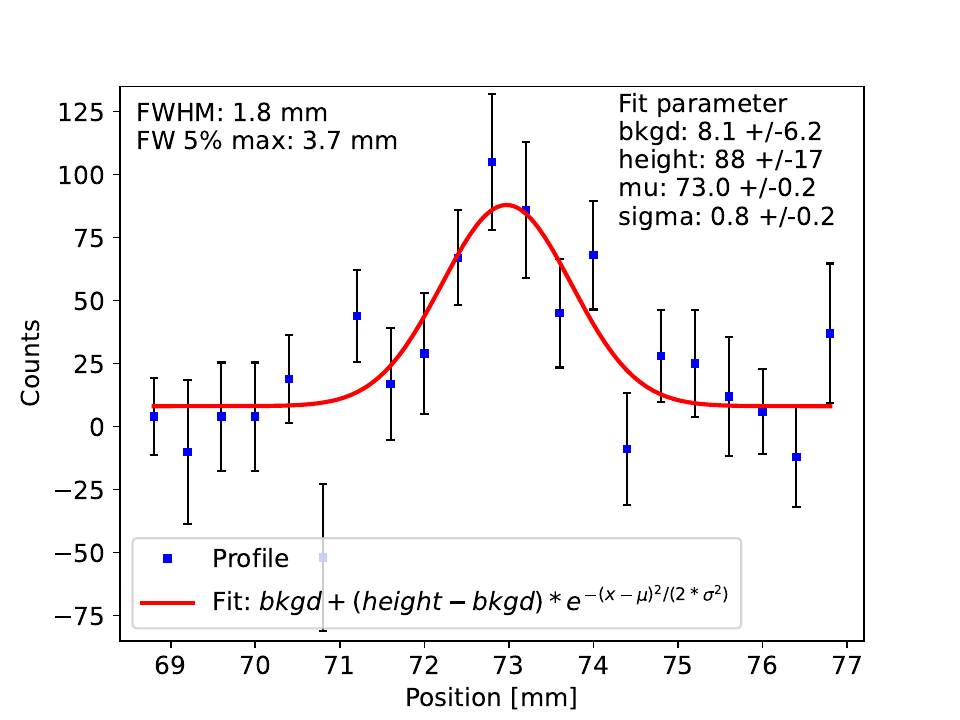}%
}
\caption{Gaussian fit of profile in x-direction of the same diffraction spot ($\omega = 55\si{\degree}$, $\gamma = 60\si{\degree}$) in the MILAND detector and the Gd-GEM. The $^{3}He$ detector acquired data for 300~s, whereas the Gd-GEM took data for 1620 s. The error bars represent the statistical uncertainty on the counts.}
\label{fig: fit_1D_diffraction spot}
\end{figure}

To investigate the shape and width of the diffraction spots, a Gaussian has been fitted to the profile in x-direction. Figure~\ref{fig: fit_1D_diffraction spot} displays the same diffraction spot ($\omega = 55\si{\degree}$, $\gamma = 60\si{\degree}$), on the left in the MILAND $^{3}He$ detector and in the Gd-GEM on the right. The MILAN $^{3}He$ detector acquired data for 300~s, whereas the Gd-GEM took data for 1620 s. Since the data of the Gd-GEM has been background subtracted, the signal to background ratio is slightly better in the Gd-GEM with 9.9 compared to 2.5 for the MILAND detector. The FWHM for the profile in the MILAND detector amounts to 3.5 mm, whereas in the Gd-GEM one gets 1.8 mm. These values cannot be compared directly, since the diffraction spots are not completely round, and the x-axis of the Gd-GEM detector was rotated 45.55$\si{\degree}$ with respect to the x-axis of the MILAND detector. The Gd-GEM detector produces neutron-induced secondary charged particles (electrons) of energies between 10 keV and 250 keV. Their different range in gas should not produce a deformation of the diffraction spots, since with the help of the uTPC method, the interaction point of the neutron on the cathode is reconstructed. For the MILAND $^{3}He$ detector, there might be a small parallax effect, since the diffracted beam is not always perpendicular to the detector surface, and the position of the neutron is calculated from the charge with the centre of gravity method.

\begin{figure}[htbp]
\centering
\subfloat[MILAND detector\label{fig: fit_2D_spot_He3}]{
\includegraphics[width=.50\textwidth]{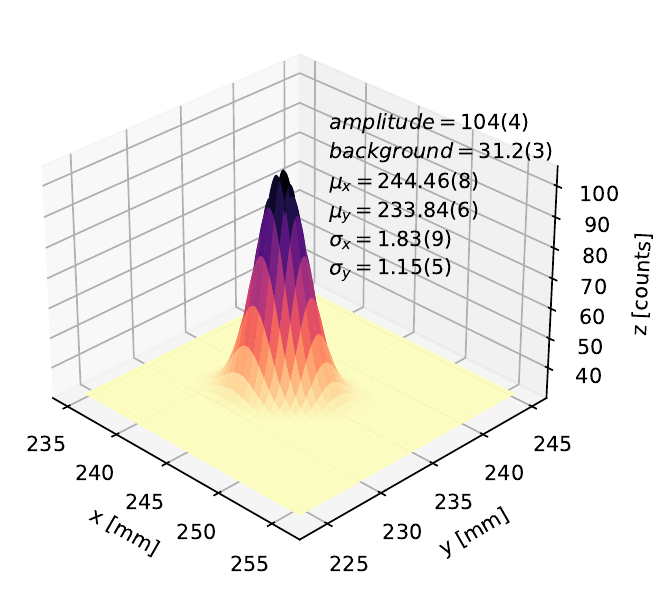}%
}
\subfloat[Gd-GEM detector\label{fig: fit_2D_spot_Gd-GEM}]{
\includegraphics[width=.50\textwidth]{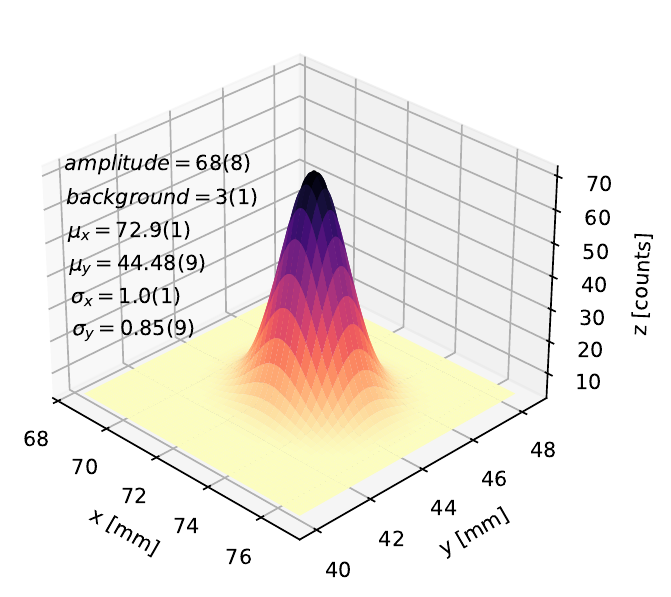}%
}
\caption{Fit of two-dimensional Gaussian to the same diffraction spot ($\omega = 55\si{\degree}$, $\gamma = 60\si{\degree}$) in the MILAND detector and the Gd-GEM. The $^{3}He$ detector acquired data for 300~s, whereas the Gd-GEM took data for 1620 s.}
\label{fig: fit_2D_diffraction spot}
\end{figure}

The fit of a two-dimensional Gaussian to the same diffraction spots is shown in figure~\ref{fig: fit_2D_diffraction spot}. The $\sigma_{x}$ of the Gaussian fit amounts to 1.83(9) mm for the $^{3}He$ detector and 1.0(1) mm for the Gd-GEM, whereas the values for $\sigma_{y}$ are 1.15(5) mm and 0.85(9) mm respectively. The 2D Gaussian of the MILAND detector is rotated with respect to the 2D Gaussian in the Gd-GEM, the $\sigma_{x}$ is thus not the $\sigma$ along the x-axis but just one of the principal axes of the Gaussian. When comparing now the two larger $\sigma_{x}$ of both detectors, the $\sigma_{x}$ of the MILAND detector is 1.8 times larger than the $\sigma_{x}$ of the Gd-GEM. For the smaller $\sigma_{y}$, the $\sigma_{y}$ of the MILAND detector is 1.4 times larger than the $\sigma_{y}$ of the Gd-GEM. The Gd-GEM was installed at 196 mm from the sample, whereas the $^{3}He$ detector was positioned at 332 mm. Assuming an equal detector resolution and ignoring parallax effects, the size of the spot should be around 1.7 times smaller in the location of the Gd-GEM. The $\sigma$ derived from the 2D fit indicates thus a similar point-spread function for both detectors. This indicates the absence of instrumental effects on the reconstruction of the diffraction spot in the Gd-GEM detector.

In the following paragraph, as a rough plausibility check, the total number of counts in the diffraction spots in the MILAN detector and the Gd-GEM are compared. The integral of the two-dimensional Gaussian fit amounts to 980.6 counts/$mm^2$ for the MILAND detector, and 359.6 counts/$(0.4 mm)^2$ for the Gd-GEM detector. Considering the different acquisition times of 300 s for the MILAND detector, and 1620 s for the Gd-GEM, one arrives at a rate of 3.3 $Hz/mm^2$ for the MILAND detector and 1.4 $Hz/mm^2$ for the Gd-GEM. According to this data analysis, the MILAND detector is thus about 2.4 times more efficient than the Gd-GEM. The Gd-GEM with the traditional x/y readout has a measured efficiency of 11.8$\%$ for neutrons with a wavelength of 2 \AA~\cite{Pfeiffer2016}. For a Gd-GEM detector with low material budget readout, as used in the measurements presented here, the detection efficiency has been simulated with Geant4~\cite{Geant4a} to be around 25$\%$ with a threshold of 0 keV at 4.51 \AA. The $^{3}He$ wire chamber has a detection efficiency of about 70 $\%$ at a wavelength at 2.5 \AA~\cite{MILAND}. Assuming a similar or higher efficiency of the $^{3}He$ wire chamber at 4.51 \AA, the efficiency calculated from the measurement of the diffraction spots slightly overestimates the efficiency of the Gd-GEM detector.

\section{Conclusion}
\label{Conclusion}
The improved prototype of the NMX detector consisting of a Triple-GEM detector with Gd converter has been successfully tested at the neutron reactor of the Budapest Neutron Centre (BNC)~\cite{BNC} and at the D16 instrument at the Institut Laue-Langevin (ILL)~\cite{ILL} in Grenoble. The measurements with Cadmium and Gadolinium masks in Budapest demonstrate that the point-spread function of the detector lacks long tails that could impede the measurement of diffraction spot intensities. This result indicates that the detector is suitable for measuring the diffraction spot intensities in crowded diffraction patterns, as detailed in reference~\cite{NMX2020}. On the D16 instrument at ILL, diffraction spots from Triose phosphate isomerase w/ 2-phosphoglycolate (PGA) inhibitor were measured both in the D16 MILAND detector and the Gd-GEM. The comparison between the two detectors shows a similar point-spread function in both detectors, and the expected efficiency ratio compared to the MILAND detector. Combined with the high rate capabilities~\cite{Thuiner2016,Pfeiffer2022} and the excellent time resolution of GEM detector~\cite{Scharenberg2020} (not evaluated during the tests described here) and lack of parallax effects, there are thus good indications that the  Gd-GEM detector fits the requirements for the NMX instrument at ESS.

\section{Annex}
\label{Annex}
This annex contains the geometrical calculations that proof that the diffraction spots on the MILAND detector and the Gd-GEM detector (as compared in chapter \ref{ILL}) come from the same reflection. 

\begin{figure}[htbp]
\centering
\subfloat[View from sample.\label{fig: detector_readout}]{
\includegraphics[width=.50\textwidth]{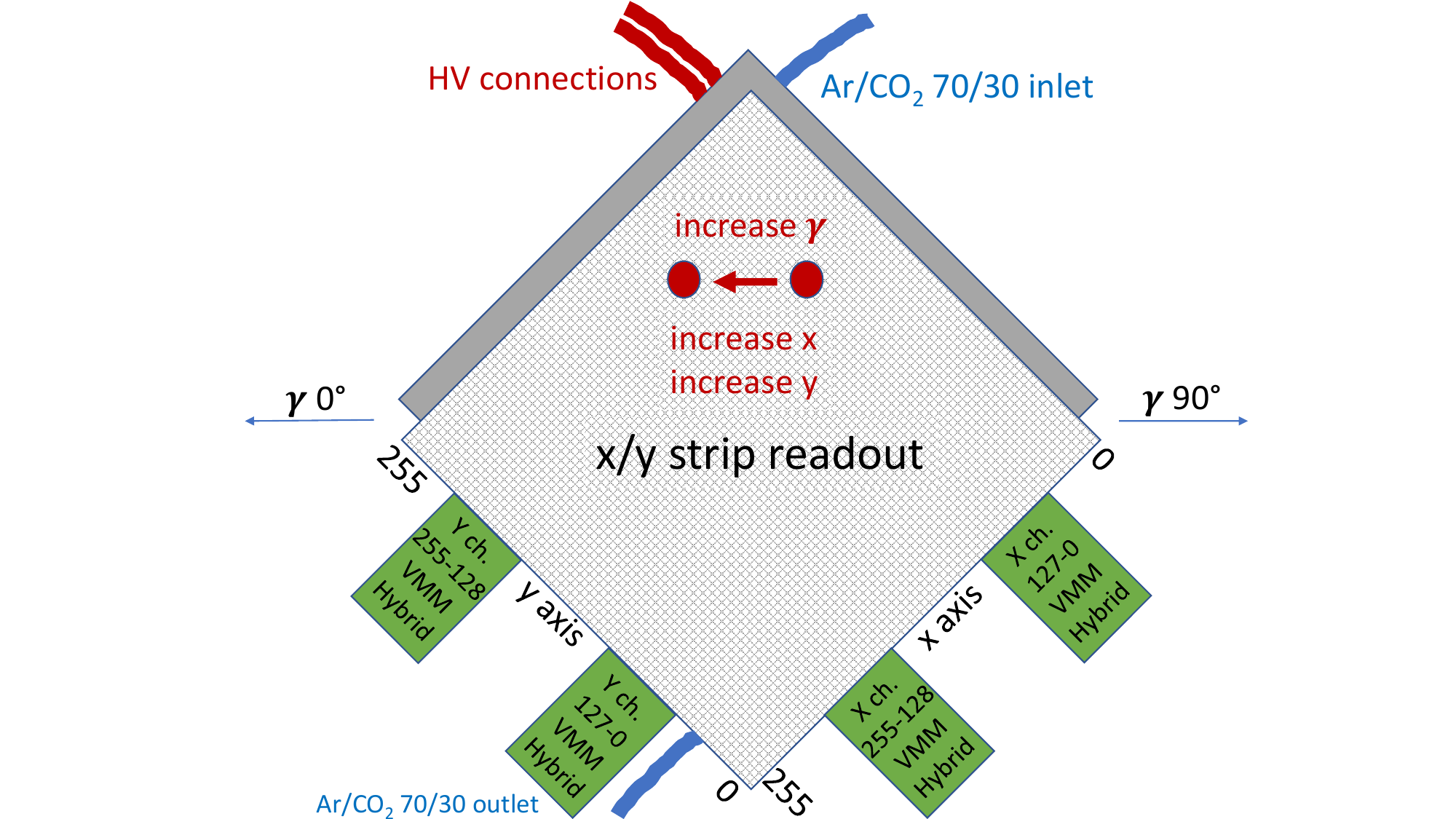}%
}
\subfloat[View from back of detector.\label{fig: detector_converter}]{
\includegraphics[width=.50\textwidth]{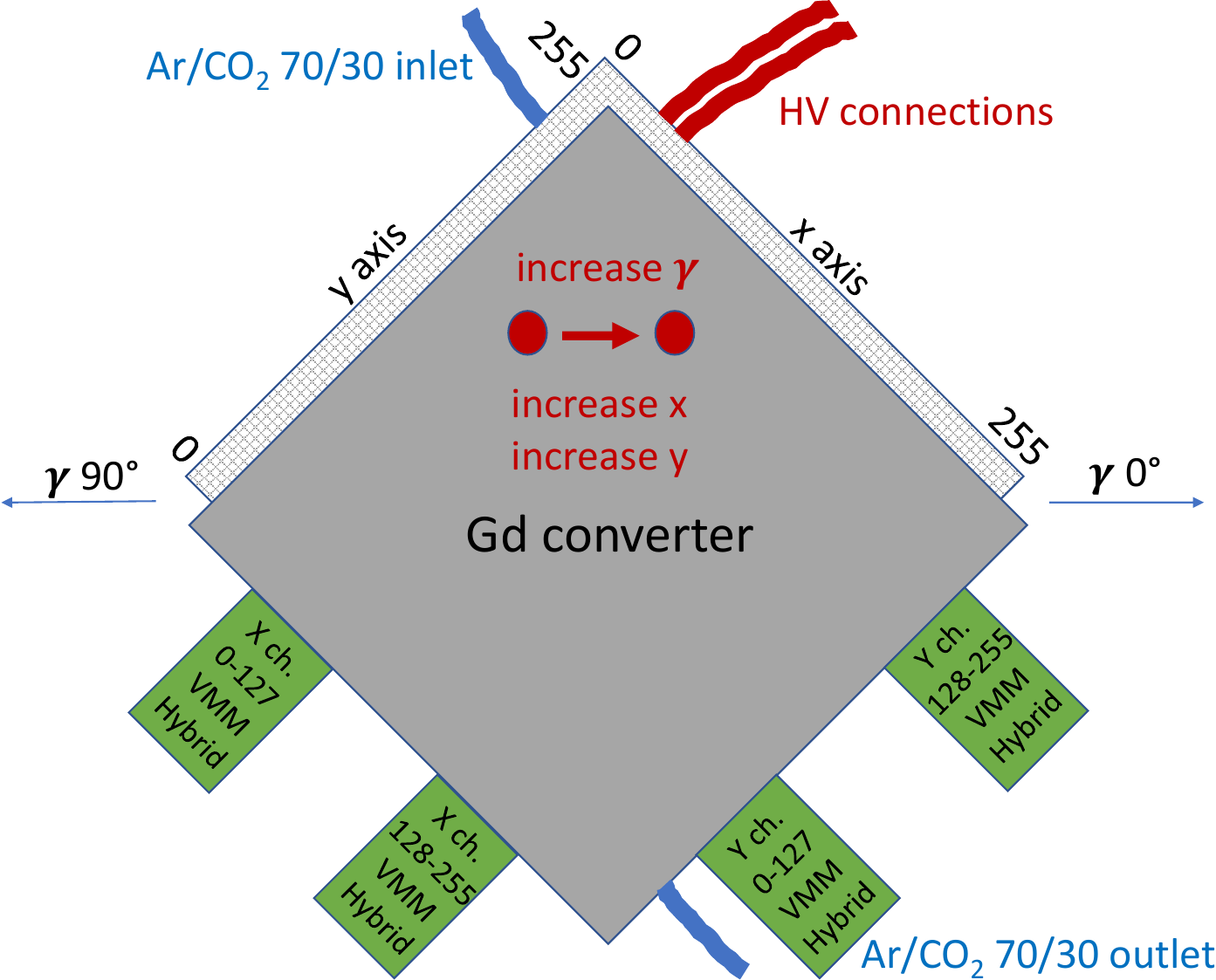}%
}
\caption{Schematic drawing of detector readout (x/y strip readout), VMM hybrids, Gd converter and movement of diffraction spots on detector depending on rotational movement $\gamma$ of detector. The neutrons coming from the sample traverse first the readout of the detector, as shown previously in figure~\ref{fig: Detector_Scheme}.}
\label{fig: drawing_detector}
\end{figure}

Figure \ref{fig: detector_readout} displays the schematic drawing of the Gd-GEM detector as it is seen from the sample. The neutrons pass first through the detector readout, then the three GEM foils and arrive finally at the cathode. If the detector rotation angle $\gamma$ is increased (detector moves to the right), the diffraction spot will move to the left towards larger x and y values. If seen from the cathode side as in figure \ref{fig: detector_converter}, an increase in $\gamma$ means that the detector moves to the left, and the diffraction spot moves to the right. To see the effect of the detector movement both in the x and in the y coordinate, the detector was installed with an angle $\beta$ of approximately $45\si{\degree}$ between the beamline and the x-axis.

\begin{figure}[htbp]
\centering
\subfloat[Shift of diffraction spots \label{fig: point_shift}]{
\includegraphics[width=.32\textwidth]{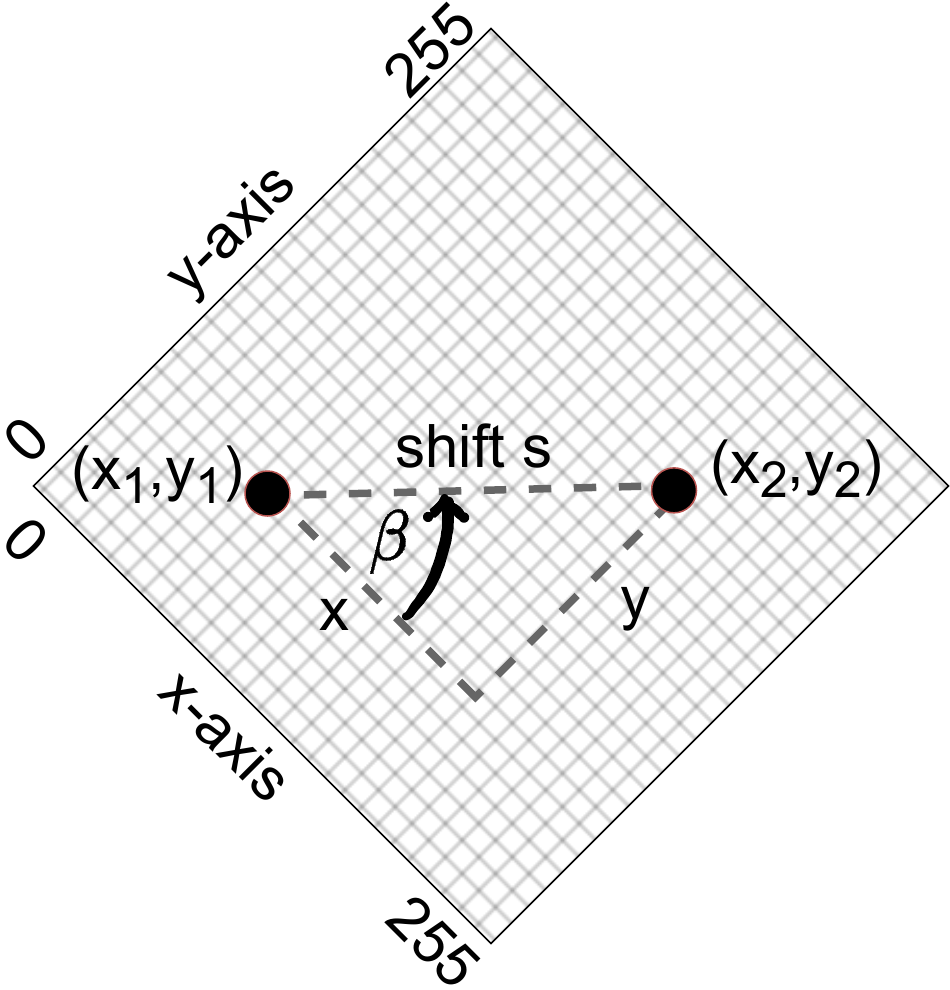}%
}
\subfloat[Rotation of detector\label{fig: angle_detector}]{
\includegraphics[width=.36\textwidth]{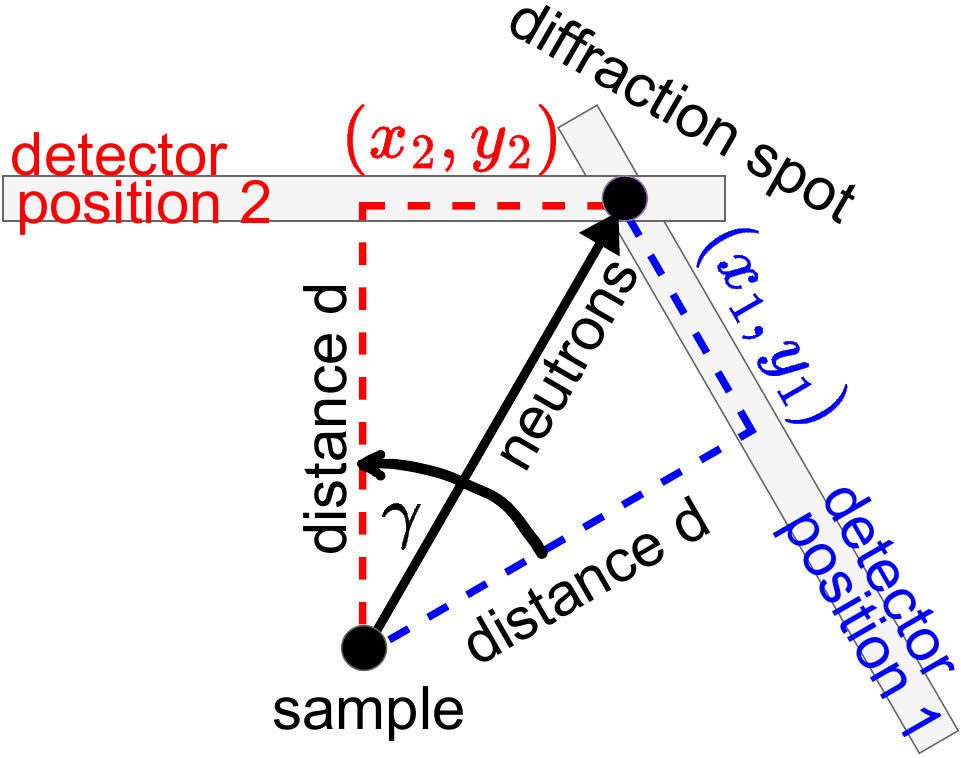}%
}
\subfloat[Rotation of sample\label{fig: angle_sample}]{
\includegraphics[width=.32\textwidth]{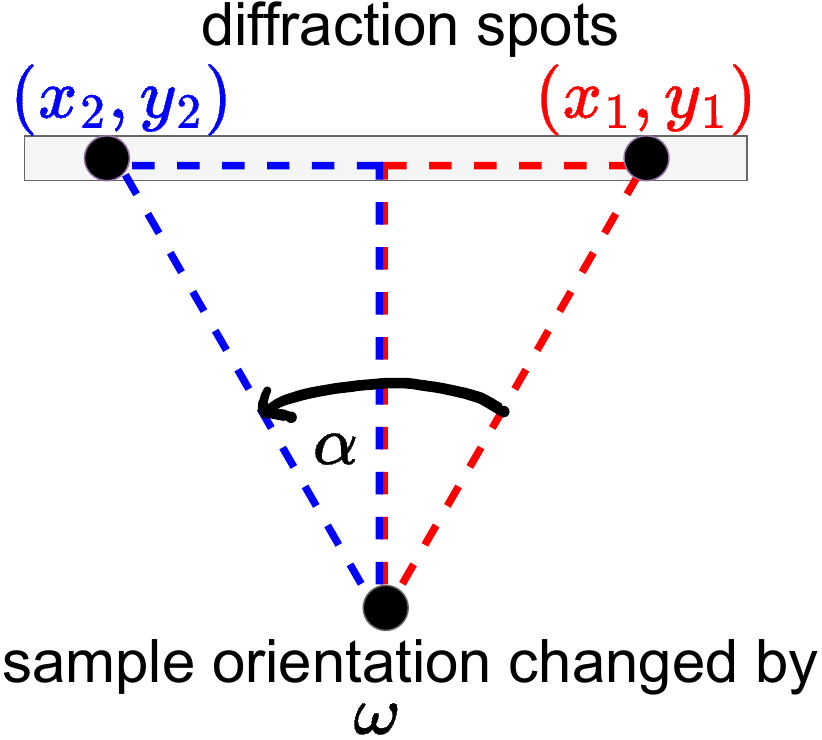}%
}
\caption{Schematic drawing of diffraction spot shift (due to detector rotation) on the detector plane(Figure \ref{fig: point_shift}). Figure \ref{fig: angle_detector} shows the effect of the detector rotation $\gamma$ on the position of the diffraction spots. Figure \ref{fig: angle_sample} displays the effect of the rotation of the sample by the angle $\omega$. This sample rotation leads to a diffraction spot shift that is equivalent to a rotation of the detector by the hypothetical angle $\alpha$. }
\label{fig: explanation_calculation}
\end{figure}

The principles of the calculations below are explained in figure \ref{fig: explanation_calculation}. A schematic drawing of the shift of a diffraction spot (due to detector rotation) on the detector plane is displayed in figure \ref{fig: point_shift}. The shift between the two spots can be calculated using the theorem of Pythagoras. The angle between the beamline and the x-axis can be determined with the Arctan (inverse Tangens) function. Figure \ref{fig: angle_detector} shows the effect of the detector rotation $\gamma$ on the position of the diffraction spots. If the distance d between the detector and the sample is known, $\gamma$ can be calculated with the Arctan function. Figure \ref{fig: angle_sample} illustrates the effect of the sample rotation by the angle $\omega$. This sample rotation leads to a shift of the diffraction spot that is equivalent to a rotation of the detector by the hypothetical angle $\alpha$. 

\begin{figure}[htbp]
\centering
\subfloat[$\gamma = 57\si{\degree}$ \label{fig: gamma_57}]{
\includegraphics[width=.50\textwidth]{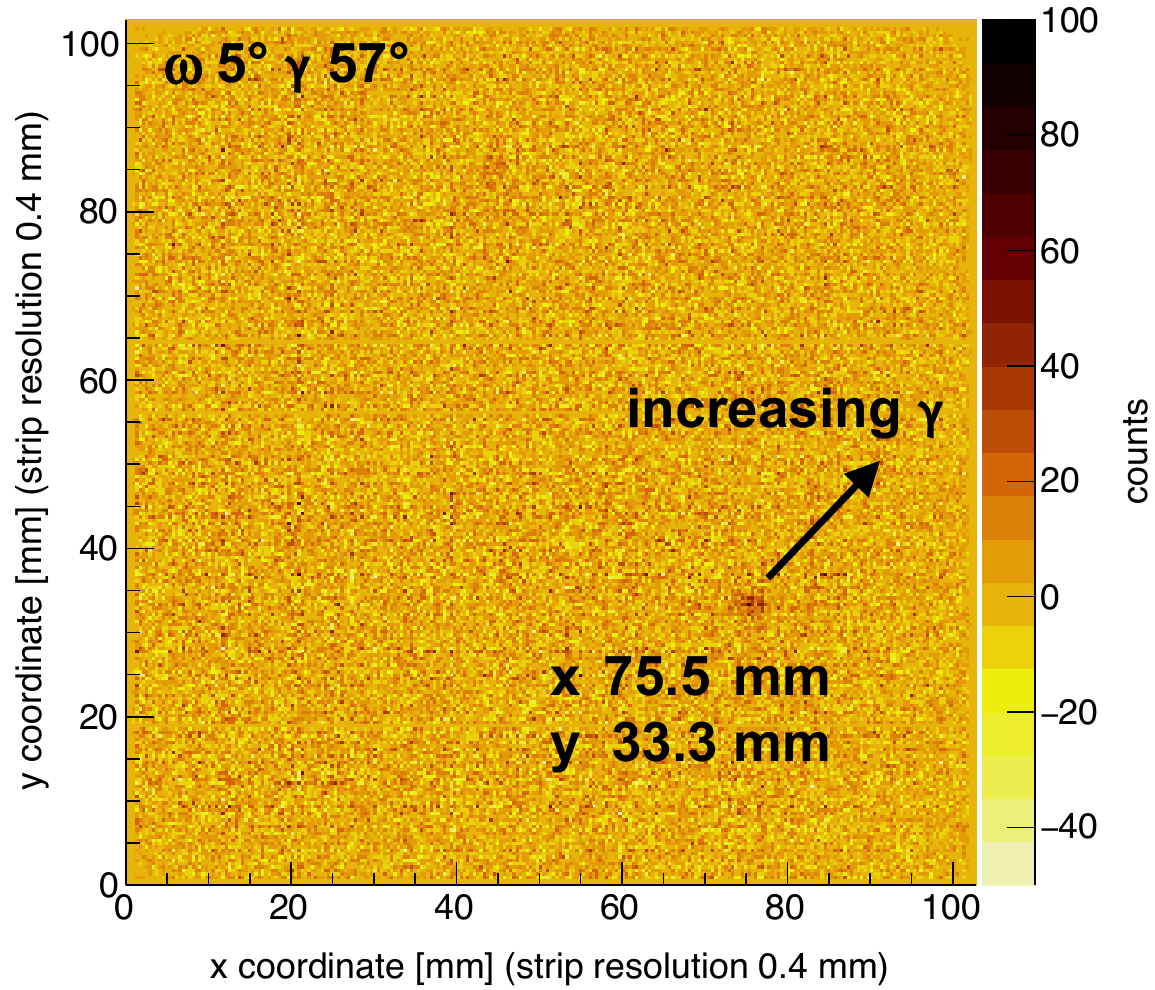}%
}
\subfloat[$\gamma = 60\si{\degree}$\label{fig: gamma_60}]{
\includegraphics[width=.50\textwidth]{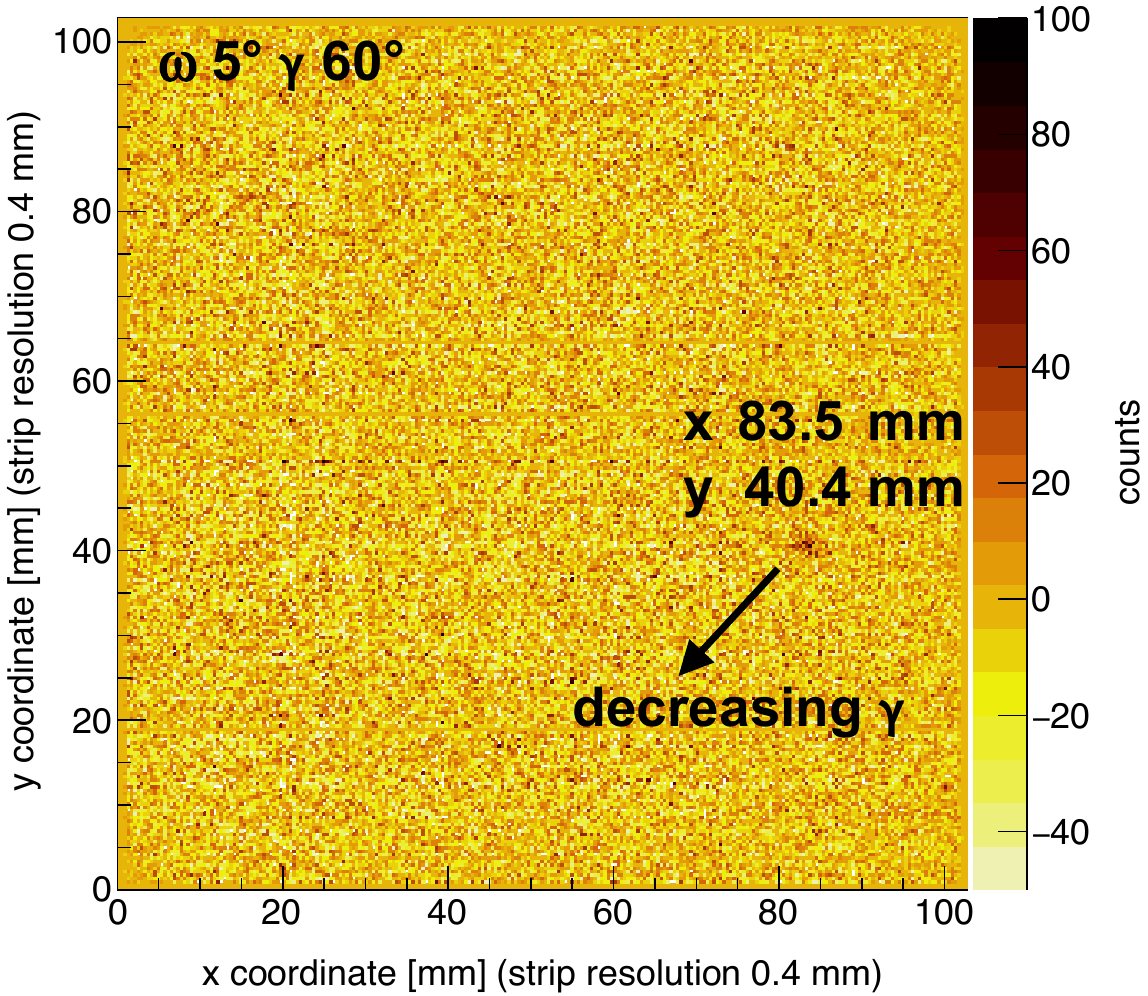}%
}
\caption{Measurement of diffraction spots from PGA inhibitor at a sample rotation of $\omega = 5\si{\degree}$. An increase in the detector rotation angle $\gamma$ moves the spot to larger x and y values.}
\label{fig: movement_spots}
\end{figure}

Figure \ref{fig: movement_spots} shows the moving of a diffraction spot from PGA inhibitor. At a sample rotation of $\omega = 5\si{\degree}$, the increase in the detector rotation angle $\gamma$ from $\gamma = 57\si{\degree}$ to  $\gamma = 60\si{\degree}$, moves the spot from the position x=75.5 mm and y=33.3 mm to x=83.5 mm and y=40.4 mm. The positions of the spot on the detector have been determined by fitting a Gaussian distribution to the data. A detector rotation of $\gamma = 3\si{\degree}$, resulted thus in a shift $s_{57\si{\degree}-60\si{\degree}}$ of the spots.

\begin{equation}
\begin{split}
s_{57\si{\degree}-60\si{\degree}} & = \sqrt{(x_{60} - x_{57})^{2} + (y_{60} - y_{57})^{2}} \\
 & = \sqrt{(83.67(26)~mm - 75.49(31)~mm)^{2} + (40.39(17)~mm - 33.28(18)~mm)^{2}} \\
 & = 10.84(47)~mm
\end{split}
\end{equation}

Using a shift $s_{57\si{\degree}-60\si{\degree}} = 10.84(47)~mm$ and a detector sample distance $d = 196~mm$, the change of the detector rotation angle $\gamma_{57\si{\degree}-60\si{\degree}}$ amounts to

\begin{equation}
\begin{split}
\gamma_{57\si{\degree}-60\si{\degree}} & = \arctan{\frac{s_{57\si{\degree}-60\si{\degree}}}{d}} \times \frac{180 \si{\degree}}{\pi} \\
 & = \arctan{\frac{10.84(47)~mm}{196~mm}} \times \frac{180\si{\degree}}{\pi} \\ 
 & = 3.17(14) \si{\degree}
\end{split}
\end{equation}

The actual angle $\beta_{57\si{\degree}-60\si{\degree}}$ between the beamline and the x-axis of the detector be calculated with
\begin{equation}
\begin{split}
\beta_{57\si{\degree}-60\si{\degree}} & = \arctan{\frac{y_{60} - y_{57}}{x_{60} - x_{57}}} \times \frac{180 \si{\degree}}{\pi} \\
 & = \arctan{\frac{40.39(17)~mm - 33.28(18)~mm}{83.67(26)~mm - 75.49(31)~mm}} \times \frac{180 \si{\degree}}{\pi} \\
 & = \arctan{\frac{7.11(25)~mm}{8.18(40)~mm}} \times \frac{180 \si{\degree}}{\pi} \\
 & = 40.99(171) \si{\degree}
\end{split}
\end{equation}

Measurements with a sample rotation of $\omega = 55\si{\degree}$ and a change of the detector rotation angle $\gamma = 60\si{\degree}$ to  $\gamma = 62\si{\degree}$, resulted in a shift $s_{60\si{\degree}-62\si{\degree}} = 6.67(42) mm$, a calculated detector rotation angle $\gamma_{60\si{\degree}-62\si{\degree}} = 1.95(12) \si{\degree}$ and an angle $\beta_{60\si{\degree}-62\si{\degree}} = 45.55(258) \si{\degree}$ between the beamline and the x-axis of the detector. All rotational shifts align thus with the expectations, which proves that there is no systematic bias in position reconstruction. Table \ref{table:GD-GEM detector_rotation} summarizes these findings.

\begin{table}[h!]
\centering
\begin{center}
\begin{tabular}{||c | c | c | c | c ||} 
 \hline
 \multicolumn{2}{||c|}{Applied}&\multicolumn{3}{c||}{Measured/Calculated} \\
 Sample rotation & Detector rotation & Spot distance & Detector rotation & $\beta_{beamline~x-axis}$ \\ [0.5ex] 
 \hline\hline
 $\omega = 5\si{\degree}$ & $\gamma = 57$ to $60\si{\degree}$ & 10.84(47)~mm & 3.17(14) $\si{\degree}$ & 40.99(171) $\si{\degree}$ \\
 $\omega = 55\si{\degree}$ & $\gamma = 60$ to $62\si{\degree}$ & 6.67(42)~mm & 1.95(12) $\si{\degree}$ & 45.55(258) $\si{\degree}$ \\
 \hline
\end{tabular}
\end{center}
\caption{Comparison of applied detector rotation and detector rotation calculated from the distance between diffraction spots.}
\label{table:GD-GEM detector_rotation}
\end{table}

For the comparison between the MILAND detector with the Gd-GEM detector, the sample orientation was changed from $\omega = 5\si{\degree}$ to $\omega = 55\si{\degree}$. In the Gd-GEM detector, the change in sample rotation from $\omega = 5\si{\degree}$ to $\omega = 55\si{\degree}$ shifted the diffraction spot by

\begin{equation}
\begin{split}
s_{Gd-GEM} & = \sqrt{(x_{55} - x_{5})^{2} + (y_{55} - y_{5})^{2}} \\
 & = \sqrt{(83.67(26)~mm - 72.97(16)~mm)^{2} + (40.39(17)~mm - 44.49(24)~mm)^{2}} \\
 & = 11.46(39)~mm
\end{split}
\end{equation}

This shift of $s_{Gd-GEM}=11.46(39)~mm$, caused by the rotation of the sample, is equivalent to a hypothetical rotation $\alpha_{Gd-GEM}$ of the detector of
 
\begin{equation}
\begin{split}
\alpha_{Gd-GEM} & = \arctan{\frac{s_{Gd-GEM}}{d}} \times \frac{180 \si{\degree}}{\pi} \\
 & = \arctan{\frac{11.46(39)~mm}{196~mm}} \times \frac{180\si{\degree}}{\pi} \\ 
 & = 3.35(12) \si{\degree}
\end{split}
\end{equation}

On the imaging plane of the MILAND $^{3}He$-detector, installed at 332~mm from the sample, the same change in sample orientation shifted the diffraction spot by

\begin{equation}
\begin{split}
s_{He_3} & = \sqrt{(x_{55} - x_{5})^{2} + (y_{55} - y_{5})^{2}} \\
 & = \sqrt{(244.38(13)~mm - 244.89(20)~mm)^{2} + (233.83(11)~mm - 213.21(11)~mm)^{2}} \\
 & = 20.63(17)~mm
\end{split}
\end{equation}

The hypothetical rotation $\alpha_{^{3}He}$ of the detector, equivalent to this shift of $s_{^{3}He}=20.63(17)~mm$, amounts to

\begin{equation}
\begin{split}
\alpha_{He_3} & = \arctan{\frac{s_{Gd-GEM}}{d}} \times \frac{180 \si{\degree}}{\pi} \\
 & = \arctan{\frac{20.63(17)~mm}{332~mm}} \times \frac{180\si{\degree}}{\pi} \\ 
 & = 3.56(3) \si{\degree}
\end{split}
\end{equation}

Since the two detector measurements were taken under the same conditions using the same detector and sample rotation, the angles $\alpha$ are as expected identical within errors.

\acknowledgments
This work was partially funded by the EU Horizon 2020 framework, BrightnESS project 676548. The authors would like to acknowledge and thank BNC for providing the use of their facilities and beam time on the ATHOS instrument. Equally, they would like to thank the ILL for providing use of their facilities and beam time on the D16 instruments.

\newpage

\bibliographystyle{IEEEtran}   
\bibliography{Performance}

\end{document}